\documentclass[manuscript,screen]{acmart}
\usepackage{listings}
\usepackage[many]{tcolorbox}    	% for COLORED BOXES (tikz and xcolor included)
\usepackage{tabularx}
\usepackage{svg}
\usepackage{amsmath}
\usepackage{multirow}
\usepackage{graphicx}
\usepackage{subcaption}
\usepackage{algorithm}
\usepackage{algpseudocode}
\newtcolorbox{boxA}{
    fontupper = \bf,
    boxrule = 1.5pt,
    colframe = black % frame color
}

\AtBeginDocument{%
  \providecommand\BibTeX{{%
    \normalfont B\kern-0.5em{\scshape i\kern-0.25em b}\kern-0.8em\TeX}}}

\acmDOI{}

\acmJournal{JACM}
\begin{document}
\setcopyright{cc}
\setcctype{by}
\acmJournal{TOSEM}
\acmYear{2026} \acmVolume{1} \acmNumber{1} \acmArticle{}
\acmMonth{1} \acmDOI{10.1145/3820772}

\title{Enhancing Software Maintenance: A Learning to Rank Approach for Co-changed Method Identification}

\author{Yiping Jia}
\email{yiping.jia@queensu.ca}
\affiliation{%
  \institution{Queen's University}
  \city{Kingston}
  \state{Ontario}
  \country{Canada}
}

\author{Safwat Hassan}
\affiliation{%
  \institution{University of Toronto}
  \city{Toronto}
  \state{Ontario}
  \country{Canada}
}
\email{safwat.hassan@utoronto.ca}

\author{Ying Zou}
\affiliation{%
  \institution{Queen’s University}
  \city{Kingston}
  \state{Ontario}
  \country{Canada}
}
\email{ying.zou@queensu.ca}

\begin{abstract}
With the growing complexity of large-scale software systems, it is challenging to accurately identify all the needed modifications to implement a specific change. 
Co-changed methods in software engineering refer to methods that frequently change together over time. 
They are often considered to be closely related, as changes made to one method may also impact the other one. 
Identifying the \textit{co-change relationships} between methods can help development teams better understand and maintain their systems (e.g., accurately identifying all needed modifications to implement a specific change). 
Prior work faces several limitations in identifying co-changed methods, e.g., generating large result sets with high false positive rates.
Focusing on the pull request (PR) level, rather than individual commits, offers a more comprehensive view of related changes that may span multiple commits, capturing essential co-change relationships.
To address the limitations of existing methods, we propose a learning-to-rank (LtR) approach that combines source code characteristics with code change history to predict and rank likely co-changes at the PR level.

Our extensive experiments, conducted on 150 open-source Java projects totaling 41.5 million lines of code and 634,216 pull requests, show that the Random Forest (RF) model outperforms other LtR models by 2.5\%–12.8\% in Normalized Discounted Cumulative Gain (NDCG@5). It also surpasses baseline methods—including support ranking, file proximity, code clone detection, FCP2Vec, and the StarCoder 2 model—by 4.7\%–573.5\% in NDCG@5.
Models trained on longer historical data (90–180 days) deliver more consistent performance, while prediction accuracy begins to decline after 60 days, suggesting a need for bi-monthly retraining. This approach offers a practical tool for software teams to prioritize co-changed methods, enhancing their ability to manage complex dependencies and maintain software quality.

\end{abstract}




\maketitle

\section{Introduction}
\begin{figure}[tb]
  \centering
  \includegraphics[width=\linewidth]{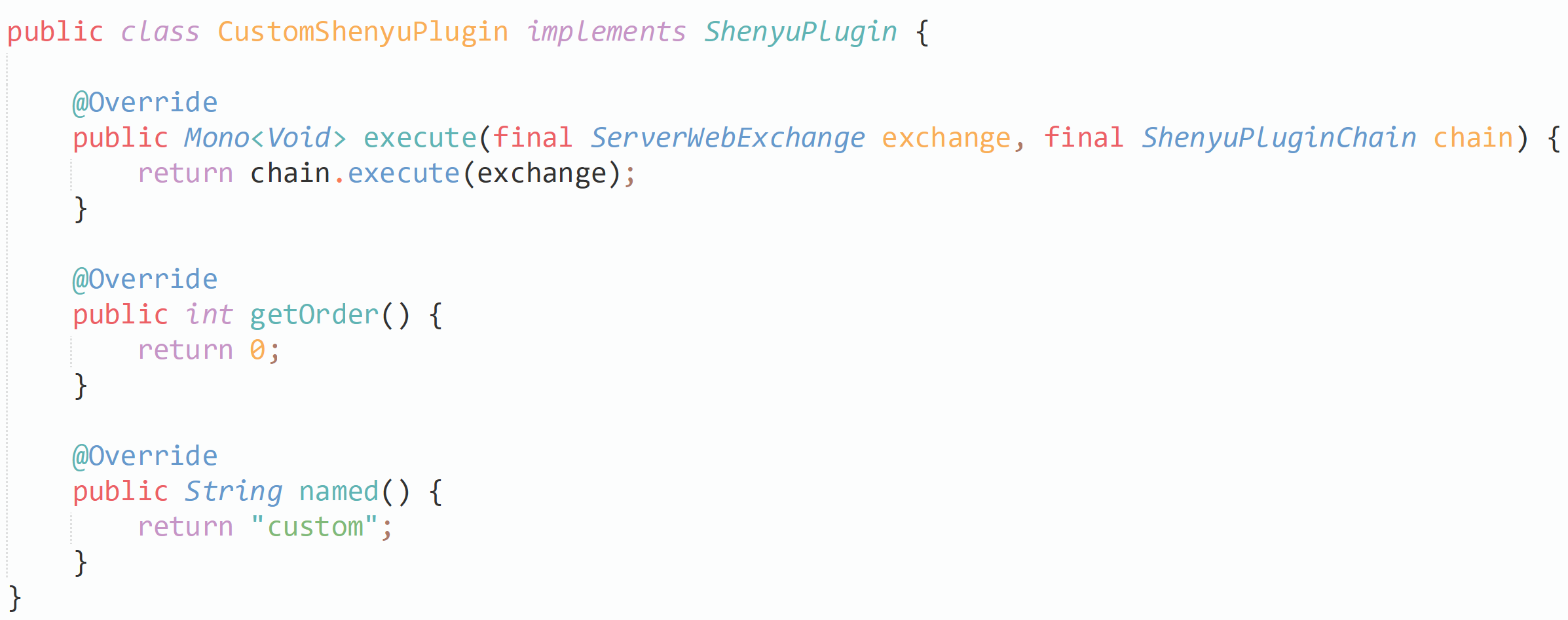}
  \includegraphics[width=\linewidth]{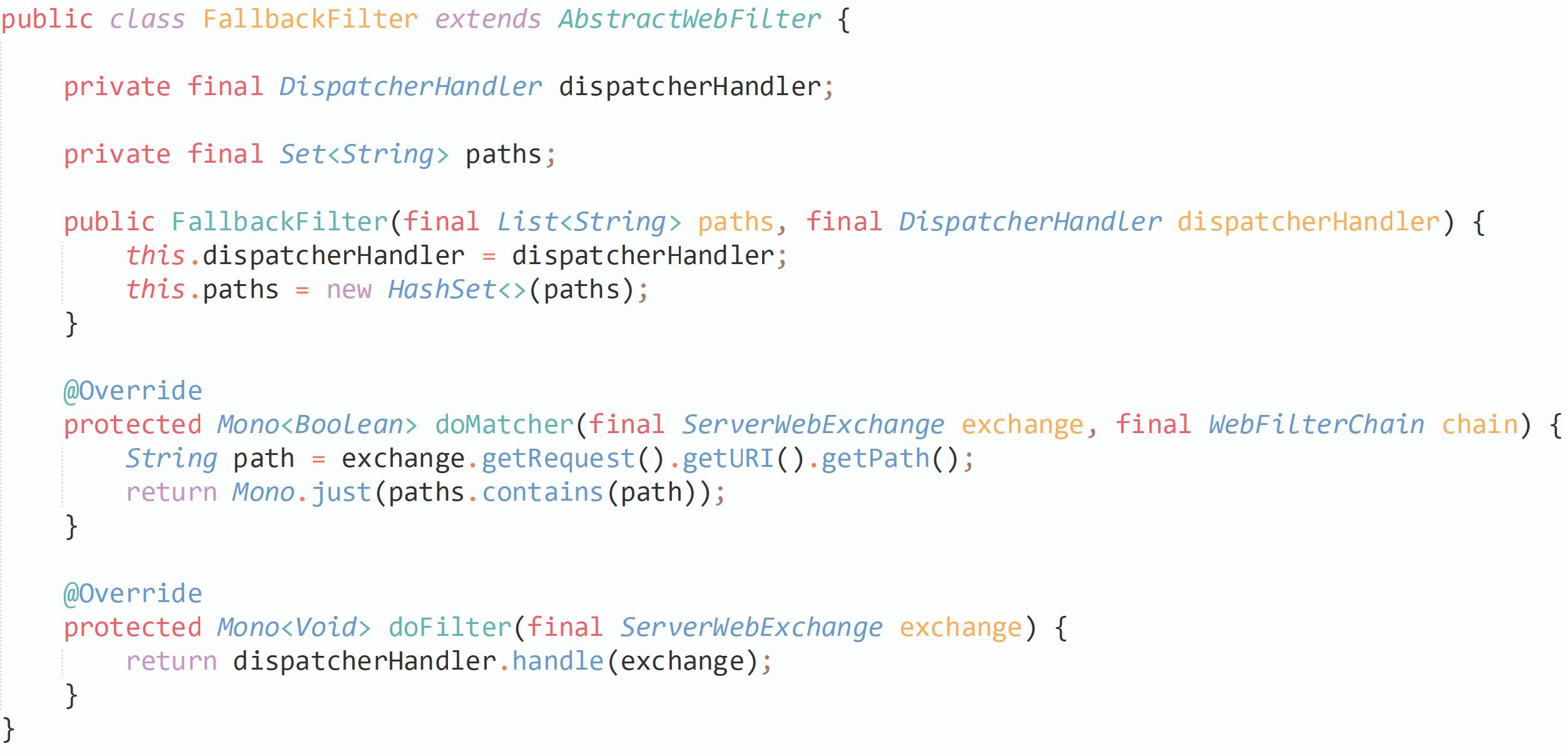}
  \caption{An example of the co-change relationship between the method \textit{`execute'} in class \textit{`CustomShenyuPlugins'} and the method \textit{`doFilter'} in class \textit{`FallbackFilter'} in the \textit{`Shenyu'} project.}
  \label{img:example}
\end{figure}

Software systems undergo continuous changes, such as adding new features, fixing defects, and optimizing the code for better performance and quality improvement. 
During software system evolution, some artifacts in the same software may frequently change together. 
For example, defect-fixing may require changes to multiple methods in different modules related to the same functionality. 
The relationship between software artifacts frequently changed simultaneously is known as the \textit{``co-change relationship''} or \textit{``evolutionary coupling''}~\cite{historank}.
%In this paper, a co-change occurrence refers to the case where two methods are changed together in a single pull request (PR). A co-change relationship, on the other hand, means that two methods are frequently co-changed across multiple PRs, indicating a recurring maintenance or functional linkage between them.
Different entities may be modified within the same pull request (PR) or commit, and such co-change can occur purely by coincidence. However, when the same pair of entities is repeatedly co-changed across multiple PRs, this recurring pattern suggests a strong relationship that requires them to be edited together. Such co-change relationships are not limited to entities within the same module, class, or package, and can happen even without explicit dependencies.
For example, as shown in Figure~\ref{img:example}, in project Apache/Shenyu\footnote{https://github.com/apache/shenyu}, two methods (i.e., the method \textit{execute} in Class \textit{CustomShenyuPlugins} and the method \textit{doFilter} in class \textit{FallbackFilter}) have been both edited together for three times (i.e., through three pull requests) during the project evolution. 
The two classes containing the two methods are not from the same package and do not share any superclass or interface (i.e., there are no explicit dependencies between the two methods). Therefore, their co-change relationship cannot be detected using code dependencies. 
Such co-change relationships could remain in developers' experience and are often not recorded in documentation or the source code~\cite{EC4SAR3}. 

Prior research shows that missing co-changes contribute to defects such as unsynchronized bug fixes in code clones~\cite{related2bugs1}, where a fix is applied to one clone instance but overlooked in other duplicates. This oversight can leave latent vulnerabilities unaddressed, resulting in recurring issues with the software’s functionality. In addition, studies show that the presence of a co-change relationship increases the likelihood of defects when related updates are not properly handled~\cite{related2bugs1}.
For instance, in the issue \textit{OF-3026}~\footnote{https://igniterealtime.atlassian.net/browse/OF-3026} of repository igniterealtime/Openfire, the intended behaviour during a fast reconnect is simple: the system should first remove the old connection from its main routing record, update the supporting lists that mirror this record, and then add the new connection, so every part agrees on where to deliver messages. The bug appeared because the main record was updated without updating the supporting lists simultaneously, which left some parts still pointing to the old connection while others pointed to the new one. This led to misrouted messages and sessions that looked active even though they were not. The developers fixed the problem by making these related updates occur in one coordinated step, removing the old entry, adding the new one, and refreshing the lists together. This case illustrates why detecting co-changes is critical: when multiple structures describe the same fact, reliability depends on changing all of them together, not one after another.

It is critical to detect co-change relationships as it helps developers localize the needed changes without missing any dependencies~\cite{zimmermann2005mining}. 
Since co-changed methods are likely to be impacted by the same changes, they may need to be tested together to ensure they continue to work correctly. 
In the literature, identifying co-change relationships has been applied to locating faults, software architecture recovery, and impact analysis~\cite{ EC4FaultLoc2, EC4SAR, EC4ImpactAnalysis} using various approaches, such as detecting co-changed clones~\cite{EConClone1, EConClone2, EConClone3, EConClone4}, applying association rules~\cite{EConClone4, zimmermann2005mining, EC4ImpactAnalysis}, and locating the co-changes in the same commits~\cite{EC4SAR3}.
However, the existing approaches suffer from the following limitations:

1) Detecting co-change relationships among clones overlooks the co-change relationships between non-cloned code, leading to an incomplete understanding of the evolutionary dynamics within the software systems.

2) Detecting a large set of co-change relationships using association rules often produces a set of unordered co-changed software artifacts.
 The result set is often excessively large, leading to a high incidence of false positives~\cite{historank}. 
 This inefficiency necessitates further prioritization of the result set to provide actionable insights to developers.

3) Detecting co-changes at a fine-grained commit level may miss the co-changed methods that might be changed in different commits and then pushed together or merged into the target branch. 
Moreover, different artifacts can be changed in the same commit coincidentally rather than due to a co-change relationship. 
Therefore, ranking co-changed artifacts at the PR level is more appropriate.

4) Existing ranking-based approaches often rely on heuristic scoring functions.
While effective at capturing strong evolutionary features, such heuristic rankings impose fixed importance on individual features and do not explicitly model how multiple heterogeneous features should be combined.
As a result, they provide limited flexibility in refining rankings when multiple candidates exhibit similar historical support.

To further demonstrate the importance of accurately detecting co-change relationships, we conducted a manual analysis on a randomly selected open-source project, namely, igniterealtime/Openfire\footnote{https://github.com/igniterealtime/Openfire}. We randomly sampled 50 methods from the project and, for each, identified the top 5 most frequently co-changed methods at the pull request (PR) level, resulting in 250 co-change method pairs. We then examined whether these co-changes also occurred at the commit level. Our analysis revealed that 67\% of the identified co-change pairs appeared in different commits within the same PR. 
This observation suggests that PR-level co-change captures non-trivial evolutionary coupling that would be missed at commit granularity, while requiring recurrence across PRs further filters out incidental co-edits.
These observations highlight the limitation of commit-level approaches in capturing cross-commit co-change relationships and motivate our focus on PR-level analysis.
We note that this analysis is based on a single repository and is intended as a motivating illustration; the generality of this observation is subsequently examined through our large-scale evaluation on 150 projects.

To address the aforementioned limitations, we propose CoRanker, a learning-to-rank framework that integrates heterogeneous features extracted from both source code and version control history to prioritize co-changed methods. Unlike heuristic ranking approaches that rely on fixed scoring functions, learning-to-rank enables the ranking model to learn how different features should be combined and weighted based on data, optimizing the ranking directly for recommendation quality. By framing co-change detection as a learning problem, CoRanker can adaptively refine rankings beyond historical frequency alone, particularly when multiple candidates exhibit similar co-change histories. CoRanker provides  an overall co-change detection framework, including data collection, feature extraction, model training, and evaluation. LtR model refers specifically to the learning-to-rank algorithm (e.g., RF, LambdaMART) used within CoRanker to rank candidate co-changed methods.  The LtR model is one of core components used within CoRanker.

We conduct experiments on 150 open-source Java projects obtained from GitHub with 41.5 million LOC and 634,216 pull requests. 
Our work aims to address the following research questions (RQs):

\textbf{{RQ1. What is the performance of CoRanker to rank co-changed methods?}}
We aim to evaluate the LtR models in retrieving co-change relationships.
We compare the performance of different LtR models and want to select the best-performing LtR model on the co-change ranking task.
The result indicates that RF has surpassed the other LtR models by a margin of 2.5\%--12.8\% in terms of Normalized Discounted Cumulative Gain (NDCG@5), achieving statistical significance with a p-value less than 0.05 in the Wilcoxon signed-rank test~\cite{wilcoxon}.

\textbf{RQ2. Can our LtR-based approach perform better than baselines?}
We compare CoRanker with six baselines (i.e., support ranking, file proximity, NiCad/clone, FCP2Vec, StarCoder 2-scalar, StarCoder 2-binary). The results show that the random forest LtR ranking schema of CoRanker outperforms the existing baselines by a margin of 4.7\%--573.5\% in terms of NDCG@5.

\textbf{RQ3. What are the important features in building the LtR models?}
We collect ten different features to build the random forest LtR model, but not all of the features contribute to its prediction power. 
Understanding the most critical features enables the streamlining of data collection processes by allowing for the elimination of less pertinent features. 
We analyze the importance of the collected features and find that the occurrence of the co-change history, the similarity of the file path, and the similarity of the authors of the methods play the most important roles in predicting the co-change relationships.

\textbf{RQ4. How soon can the model reach a consistent and accurate performance in identifying co-change relationships?}
We aim to find the shortest period for which the data is sufficient to train the model and produce a good performance. 
The shorter historical data is needed, the more likely such an approach can have fewer obstacles to be applied in practice. 
We find that the RF model with 90 days of history to label the training data can reach decent prediction power (i.e., the highest measured NDCG@5 being 0.88), and there is no significant difference in its prediction power compared to the RF model with 180 days of history to label the training data.
We also noticed that the model with 30 days of history to label the training data shows significantly lower performance than the models with 180 days of training data by a margin of 20.7\% when tested on predicting co-changed methods for a 270-day period.

Our work is characterized by the following key contributions:
 
1) We propose a learning-to-rank framework that learns how to combine historical, structural, and semantic features to prioritize co-changed method pairs, rather than relying on fixed heuristic scoring functions.

2) We evaluate CoRanker on a large-scale dataset of 150 practical projects and make the replication package available\footnote{https://github.com/jia-yp/co-change-replication}. 

3) CoRanker can effectively predict the co-change methods and outperform the baselines by a margin of 4.7\%--573.5\% in terms of NDCG@5.

4) We provide large-scale empirical insights into the relative importance of historical, structural, and semantic features for co-change ranking, showing that while historical co-change frequency dominates, learning-based integration of additional features substantially improves early-rank recommendation quality.

5) Our work provides insights into the deployment of CoRanker (e.g., the needed re-training and testing periods to keep the prediction accurate for current data) in a practical environment.    
    
6) We explore the possibility of using Large Language Models (LLMs) for co-change prediction. We gain insights into how pre-trained code models may struggle to match our LtR approach.

The rest of the paper is organized as follows.
Section \ref{sec:bg} introduces the background of LtR algorithms.
Section \ref{sec:expSetup} describes CoRanker and the experiment setup.
Section \ref{sec:results} presents the results of our research questions.
Section \ref{sec:implication} explores the implications of CoRanker.
Section \ref{sec:threat} discusses the threats to the validity of our study.
Section \ref{sec:related} summarizes the prior studies and related work in the co-change detection domain.
Section \ref{sec:conclusion} concludes our work and discuss the possible future work.

\section{Learning to Rank (LtR) Algorithms}
\label{sec:bg}
LtR is a Machine Learning (ML) approach designed to train models for ranking items according to their relevance to a query. 
Unlike traditional classification or regression techniques, which predict independent labels or values, LtR focuses on arranging items, aiming to present the most relevant results to users by taking account of multiple features to derive a ranking schema. 
LtR is prominent in search-based applications, such as search engines and recommendation systems.
LtR helps us recommend only the most relevant items, reducing the likelihood of overwhelming developers with an extensive list of candidates. This approach minimizes distractions and false positives, ensuring developers focus on the most pertinent suggestions.
In our study, we employ the LtR algorithms to identify software methods that are most likely to undergo co-changes with a given query method.

The ranking model is trained on a dataset of query-candidate pairs, with the label of the relevance of each candidate to the corresponding query. 
The model can then make predictions on new query-candidate pairs and rank the candidates based on their predicted relevance.
Different models have different mechanisms to rank the candidates. 

However, they can be generally categorized into the following three categories:

\noindent \textbf{1) Point-wise ranking:} These models predict the relevance score of a specific query-candidate pair independently, treating the problem as a regression or classification task.
\textit{Linear Regression (LR)} assumes a linear relationship between input features and the output score. It is simple, fast, and interpretable, but limited in capturing non-linear patterns.\\
\textit{Multiple Additive Regression Trees (MART)} is a boosting-based model that builds an ensemble of trees, each correcting the prediction errors of the previous ones. We trained MART with 1000 trees, 10 leaves per tree, and a learning rate (shrinkage) of 0.1.\
\textit{Random Forest (RF)} constructs multiple bags of decision trees using bootstrapped data samples and random subsets of features. Our configuration uses 300 bags, each containing 1 tree with 100 leaves. The feature sampling rate is 0.3, and the learning rate is 0.1.\
\textit{Deep Neural Network (DNN)} models complex, non-linear relationships using layered representations. We used a single hidden layer with 10 nodes, trained over 100 epochs with a learning rate of 0.00005.

\noindent \textbf{2) Pair-wise ranking:} These models compare the relevance of two candidates for a given query and learn which one should be ranked higher. This is formulated as a binary classification task over pairs.
\textit{RankNet} uses a neural network to predict which of two candidates is more relevant, based on their feature representations. The model was trained with 1 hidden layer, 10 nodes, 100 training epochs, and a learning rate of 0.00005.\
\textit{LambdaMART} integrates boosting with ranking-specific objectives by adjusting the gradients based on how much a candidate swap would affect metrics like NDCG. We trained LambdaMART with 1000 trees, 10 leaves per tree, and a learning rate of 0.1.

\noindent \textbf{3) List-wise ranking:} These models evaluate and optimize the order of a full list of candidates simultaneously, directly aiming to improve list-level metrics.
\textit{Coordinate Ascent} iteratively adjusts one feature weight at a time to maximize a ranking metric such as NDCG. We used 5 random restarts, 25 iterations per dimension, and a performance tolerance of 0.001.\
\textit{ListNet} uses a probability distribution over permutations to model list rankings, then minimizes the distance between predicted and ideal rankings. We trained ListNet over 100 epochs.\
\textit{AdaRank} is a boosting method that prioritizes training examples that are poorly ranked in earlier iterations. We used 500 training rounds, but due to observed convergence issues in multiple projects, we excluded AdaRank from our final analysis.

This range of models allows us to compare ranking strategies that vary in complexity and objective formulation, ultimately helping us identify the most effective approach for co-change prediction.

\begin{comment}
    
However, they can be generally categorized into the following three categories:

\noindent \textbf{1) Point-wise ranking:} These models predict the relevance of a specific query-candidate pair and rank the candidates according to the relevance score. They typically use a supervised learning approach and can be trained using a variety of algorithms, such as logistic regression or neural networks. 

\noindent \textbf{2) Pair-wise ranking:} These models compare the relevance of two candidates for a given query. They usually use a binary classification model to select a better one from a pair of candidates. When ranking multiple candidates, the model compares all the combination of candidate pairs and rank the candidates by the number of times a candidate is considered better.

\noindent \textbf{3) List-wise ranking:} These models rank a list of candidates for a given query. They typically use a probabilistic approach, such as the ListNet~\cite{listnet} or ListMLE~\cite{listmle} algorithm, to model the entire list of candidates at once. The models take the entire candidate list as input and output all the scores in a list.
\end{comment}

In our research, we investigate nine models that are popularly used in Software Engineering work~\cite{widelyused1, widelyused2}. 
The models are listed in Table~\ref{tab:models}, which consists of four point-wise models, two pair-wise models, and three list-wise models.
In our experiment, AdaRank struggles with convergence particularly when optimizing for non-linear measures like NDCG~\cite{adarank, LtR-ndcg, LtR-direct}. 
We decide to omit AdaRank from our analysis. 

\begin{table}[tb]
  \caption{The adopted models in our research.}
  \label{tab:models}
  \begin{tabular}{ll}
    \toprule
type                        & model             \\
    \midrule
Point-wise & Linear Regression (LR)~\cite{LinearRegression} \\
Point-wise & Multiple Additive Regression Trees (MART)~\cite{MART}      \\
Point-wise & Random Forest (RF)~\cite{randomforest}    \\
Point-wise & Deep Neural Network (DNN)    \\
Pair-wise  & RankNet~\cite{ranknet}          \\
Pair-wise  & LambdaMART~\cite{lambdamart}    \\
List-wise  & Coordinate Ascent~\cite{CoordA} \\
List-wise  & AdaRank~\cite{adarank} \\
List-wise  & ListNet~\cite{listnet}  \\
  \bottomrule
\end{tabular}
\end{table}

\section{Experiment Setup}
\label{sec:expSetup}
\begin{figure*}
  \centering
  \includegraphics[width=\textwidth]{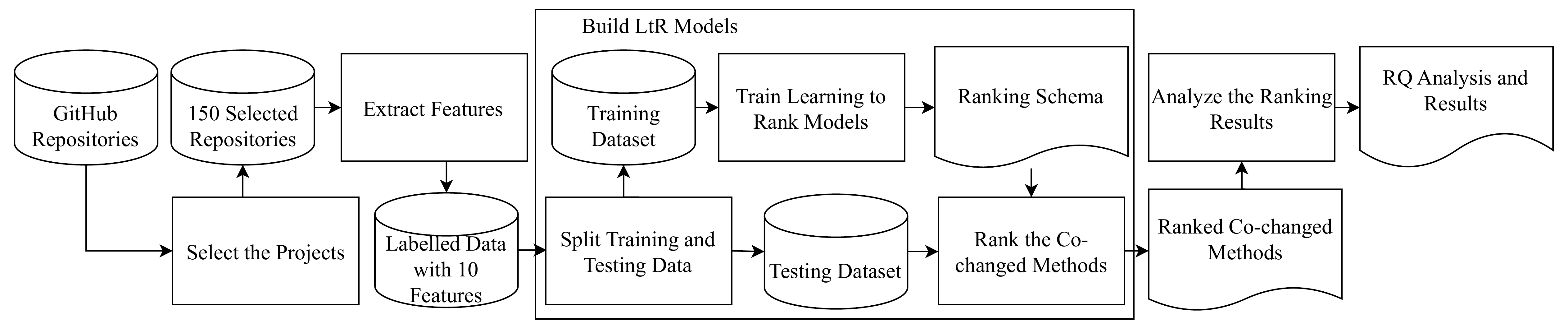}
  %\includesvg[width=\textwidth]{img/flowchart.svg}
  \caption{An overview of CoRanker.}
  \label{img:approach}
\end{figure*}

Figure~\ref{img:approach} gives an overview of CoRanker. 
We select 150 open-source Java projects from GitHub following established project selection criteria to ensure the generality of CoRanker. 
Many older or widely used co-change datasets such as git2net~\cite{git2net} and CodeChangeNet~\cite{codechangenet} lack the precise information we require, such as method-level breakdowns and PR-level co-change history. CoRanker specifically relies on PR-level labels and method-level changes to build a more accurate representation of co-change relationships. These details are often absent in standard commit-level datasets, which would prevent us from fully applying or evaluating CoRanker. Instead, we selected real-world Java projects that include the PR and method-level data necessary for our study, and we publish our dataset in the replication package.
We build a golden dataset for co-change methods based on the number of co-changes between method pairs in history. 
The golden dataset is subsequently split into training and testing sets for training LtR models. 
Once trained, these models establish a ranking schema that orders the co-changed methods based on the predicted likelihood of co-changing. 
The efficacy of the ranking schema is assessed through the analysis of the ranking results to answer the RQs. 

\subsection{Project selection}
We apply the following criteria to select our projects:

\noindent \textbf{1) Open-source Java projects:} We examine all open-source Java projects hosted on GitHub. Targeting one programming language and one platform can facilitate our data collection process and make the evaluation result more consistent. 
Since Java is one of the most popular programming languages in software development ~\cite{JavaPopular} and there are various tools to support the data collection process, we use Java projects as our primary selection. 
Although we study mainly Java projects, we believe that CoRanker can be applied to other programming languages.
    
\noindent \textbf{2) Active repositories:} 
For our analysis, we rely on a substantial period of historical data to accurately label and assess the co-change behaviour. 
Therefore, we prioritize projects that offer an extensive archive of historical data, and select those that have been active for over two years and updated in the last six months. 
This criterion helps us identify projects with a depth of information that potentially meets our research requirements.
    
\noindent \textbf{3) Repositories with merged pull requests:}  
The repository has at least 200 merged pull requests to its default branch. 
The presence of at least 200 merged pull requests in a repository serves as an indicator that there is sufficient information for our analysis at the PR level. 
    
\noindent \textbf{4) Software system repositories:}
GitHub hosts many repositories that are not software systems, such as personal code collections, tutorials, coursework, or datasets. To focus on software evolution in practice, we exclude such repositories and retain only those that represent software systems. Specifically, selected repositories must contain a non-trivial codebase with a standard software structure, such as source directories and build configurations, and involve collaborative development by multiple contributors. We also manually inspect repository documentation (e.g., README files) to exclude repositories that primarily serve educational or demonstrative purposes.

Among the 36k repositories that meet the first three criteria we design, we randomly select 150 subject repositories for the experiments, which is more than the related work~\cite{historank,age-distance,AssociateRule1,zimmermann2005mining,EConClone1,EConClone2,EConClone3,EConClone4}.
We then manually examine the 150 projects to ensure the repositories meet the fourth criterion (i.e., represent actual software systems).
Any project found not meeting this criterion is replaced with another randomly selected project, which then undergoes the same examination.
Descriptive statistics of the selected projects are listed in Table~\ref{tab:projects}, illustrating a wide range of diversity regarding the size and lifespan of the repositories.

\begin{table}[tb]
  \caption{Average and 5-number summary of different characteristics of the selected projects.}
  \label{tab:projects}
  %\resizebox{\linewidth}{!}{

  \begin{tabular}{*7l}
    \toprule
&Avg.&  Min. & 1st Qu. & Median & 3rd Qu. & Max. \\
    \midrule
\# LOC      &     276k      & 3.6K & 57.7K & 145K& 317K &1.9M\\
%\# methods & 204k & 2.4k & 21k \\
\# contributors &  175     & 3 & 81 & 152 & 267 & 4,985 \\
\# Pull Requests & 4,228 & 208      &1,091 & 2,299 & 5,248 & 30,420\\
Lifespan (years)&   7.2    & 2.2 & 5.4 &7.2  & 9.2 &13.7 \\
%\midrule
%Total \# LOC: &\multicolumn{6}{l}{ 41M}\\
  \bottomrule
\end{tabular}
%}
\end{table}

\subsection{Feature collection}
Given a query method $M_q$ and a group of candidate methods $M_C=\{M_{c1}, M_{c2}, ..., M_{cn}\}$, the LtR models use predictor features of all the query-candidate pairs $\{ \{M_q, M_{c1}\}, \{M_q, M_{c2}\}, ..., \{M_q, M_{cn}\} \}$ to predict the relevance between the query method $M_q$ and the candidates $M_C$.
In this paper, the query and the candidates are all non-test methods that are co-changed by at least one pull request in software projects. 

As we study the co-change relationship between a method pair, we suppose that one method in a project is edited and can be passed as a query to the system to identify the co-changed methods. 
We compare all the other methods in the project and select the top-ranked methods that are most likely to be edited together with the query. 
The top-ranked methods are predicted to have co-change relationships with the query method.

As shown in Figure \ref{img:approach}, we collect the predictive features from two different sources: 1) the source code of the project and 2) the edit history from the version control system.
The collected features are used as indicators to show the relationship between each method pair and serve as input predictor features for building the model.
We have designed ten features to represent the relationships of a query-candidate pair. 
The description of the ten features is presented in Table~\ref{tab:features}. 
The features are also described as follows:

\textbf{1) Historical features:} 
We use FinerGit~\cite{finergit} to collect the edit history of all the methods in the project by mining the version control system.
FinerGit takes a Git repository as input and generates another repository where every method in the original repository is transformed into a separate file. The generated repository supports all the Git commands so that we can collect the edit history with the Git-log command. The Git-log command enables us to collect the commit ID, the author, and the commit time. We then collect the pull requests associated with each commit using the official GitHub API\footnote{https://docs.github.com/en/rest}, so we can identify all methods modified by the same pull request.

Given a candidate method $M_c$ and a query method $M_q$, the historical features are defined as follows:

\textit{\textbf{Number of co-changes:}} is the total number of pull requests where the candidate method $M_c$ is co-changed  with the query method $M_q$.
    
\textit{\textbf{Authors similarity:}} Let $A_c$ and $A_q$ represent the sets of authors contribute (i.e., created or modified) to the methods $M_c$ and $M_q$ respectively. 
The authors similarity is the Jaccard similarity between the two sets of authors $A_c$ and $A_q$.

\textbf{2) Static features:} We apply static analysis of the latest source code of a project and compute the static relevance features as follows: 

\textit{\textbf{Semantic similarity:}} This feature represents the similarity between the semantic meaning of the two methods. First, we transform the source code of a method, written in the programming language, into a high-dimensional vector with CodeBERT~\cite{codebert}. 
CodeBERT is a Transformer-based deep learning model. It takes a sequence of lexical tokens, e.g., a piece of code or a short paragraph of natural language, and transforms it into a vector that contains 764 scalar values representing the lexical and semantic meaning of the input. 
We compute the semantic similarity as the cosine similarity of both vectors transformed from the query method $M_q$ and candidate method $M_c$.

While transformer models can be resource-intensive, CodeBERT is one of the most efficient in its class and performs well on standard GPUs. In our setup, embeddings for methods are computed in under 0.1 seconds per method using an A100 GPU. To optimize runtime efficiency, embeddings can be computed offline and reused unless the method changes. However, as shown in Section~\ref{sec:results}, the impact of the semantic similarity feature on overall model performance is minimal—removing it reduced NDCG@5 only slightly from 0.8394 to 0.8386. Therefore, although CodeBERT offers deeper semantic insights, it is not essential for high performance and can be omitted in resource-constrained settings. We include it primarily as an exploratory feature to assess the potential benefits of embedding-based representations in our ranking model.
    
\textit{\textbf{Path similarity:}} This feature represents the similarity between the query and candidate methods' file paths. We first split the file paths with the separator ``/'' and obtain a sequence of tokens. We then compute the similarity between the two paths $P_1$ and $P_2$ as follows:
    $$Similarity_{(P_1,\,P_2)} = \frac{number\, of\, total\, common\, tokens_{(P_1,\,P_2)}}{max(length\, of\, paths_{(P_1,\,P_2)})}$$
For example, given the query method with file path \textit{``a/b/c''} and the candidate with file path \textit{``a/d''}, the similarity between the two file paths is 1/3 because the two paths have one common token, and the maximum length of the paths is three.
    
\textit{\textbf{Code dependency:}} This feature represents the calling relationships between the query and candidate methods. We mine the dependency with a tool named Scitool Understand\footnote{https://www.scitools.com/}. The value of the feature is the sum of the number of times that the query method and the candidate method call each other.
    
\textit{\textbf{Hierarchy similarity:}} 
This feature is a boolean value, and it is true when the methods $M_q$ and $M_c$ share a common super-class; otherwise, the value is false. 
    
\textit{\textbf{Code clone:}} We use Nicad~\cite{nicad} to extract the cloned method pairs in the software repository. Nicad is a clone detector that provides reliable and accurate clone detection~\cite{EConClone1}. Nicad can detect similar code fragments in the source code and gives a similarity score between 0 and 100. We use the default setting for near-miss function clone detection, where the dissimilar threshold is set to 0.3. We use the score directly as the value of the clone feature.
    
\textit{\textbf{Package similarity:}} Similar to the file path similarity, we extract the package of the method from the source code with a regular expression and compute the similarity feature based on the query and the package of the candidate method.
    
\textit{\textbf{Argument type similarity:} }This feature represents the similarity between the types of arguments of the query and candidate methods. The value of the feature is the Jaccard similarity ~\cite{jaccard} between the sets of types defined as the method's input.
    
\textit{\textbf{Argument name similarity:}} Similar to the computing of the argument type similarity, we compute with the Jaccard similarity ~\cite{jaccard} of the argument names.

\subsection{Correlation and redundant analysis}

The existence of highly correlated features can prevent us from further analyzing the contribution of each feature to the predictive power of the model. Therefore, we conduct a correlation analysis among all the collected features and eliminate redundant ones. 

We use the Spearman rank correlation coefficient ~\cite{spearman1961proof} to conduct correlation analysis because of its robustness against non-normally distributed data. We compute the Spearman rank correlation coefficient between each pair of features and set 0.7 as a threshold. If there is any pair of features with a correlation coefficient greater than 0.7, we consider them to be redundant and we select only one of them.
The correlation analysis shows a high correlation between \textit{``package similarity''} and \textit{``path similarity''}. 
We decide to eliminate the ``package similarity'' because it is available only in some programming languages, such as Java. Therefore, keeping the ``path similarity'' would not prevent us from generalizing CoRanker to other programming languages.
The other predictor features do not have a highly correlated relationship. 

We also perform redundant analysis using the \textit{redun()}\footnote{https://www.rdocumentation.org/packages/Hmisc/versions/5.1-1/topics/redun} function in R. The redundant analysis finds no additional redundant features beyond those identified through correlation analysis in the dataset.
\begin{table}[tb]
  \caption{Features representing the relationships between a query method $M_q$ and a candidate method $M_c$.}
  \label{tab:features}
  \begin{tabularx}{\linewidth}{>{\hsize=.8\hsize}X
  >{\hsize=.1\hsize}X
  >{\hsize=2.1\hsize}X}
    \toprule
Feature                & DT* & Description                                                           \\
    \midrule
Number of co-changes   & N & Number of times the two methods are changed together at the PR level.    \\
Authors similarity      & N & Fraction of common developers who have contributed to both methods. \\

Semantic similarity      & N& The semantic similarity between the two methods.                     \\

Path similarity          & N& The directory and file name similarity between the two methods.      \\

Code dependency          & N& Presence of dependencies, such as method calls, between the two methods.                     \\

Hierarchy similarity    & C & Whether the methods share the same superclass.                            \\

Code clone              & N & Whether the two methods are code clones.                                       \\

Package similarity       & N& The similarity between the packages of the classes.                    \\

Argument type similarity& N & The similarity between the types of the arguments.                   \\

Argument name similarity & N& The similarity between the names of the arguments.   \\
  \bottomrule 
\multicolumn{3}{l}{\textbf{* Data Type (DT):} (C) Categorical – (N) Numeric}
\end{tabularx}
\end{table}

\subsection{Data preparation}
In our study, CoRanker employs supervised LtR models, which require predefined correct answers (i.e., \textit{``labels''}) to fulfill the training process. 
These labels indicate the desired order of documents (i.e., candidate co-changed methods) based on their relevance to the query (i.e., the query method). 
We determine relevance as the frequency of co-changes between methods in software development history, as outlined in Algorithm \ref{alg:dataset}. 
Specifically, we label our data based on how often a query method $M_q$ and a candidate method $M_c$ are co-changed in the subsequent six months following the collection of features that describe them.
For example, if the query method $M_q$ co-changed five times with method \textit{a}, ten times with method \textit{b}, and three times with method \textit{c}, we label the method pairs $\{M_q, a\}$, $\{M_q, b\}$, and $\{M_q, c\}$ as 5, 10 and 3 respectively. 
The labeling process is done automatically without human involvement by extracting the historical pull request information.

\subsection{Training and testing data}
\begin{figure}[tb]
  \centering
  \includegraphics[width=\linewidth]{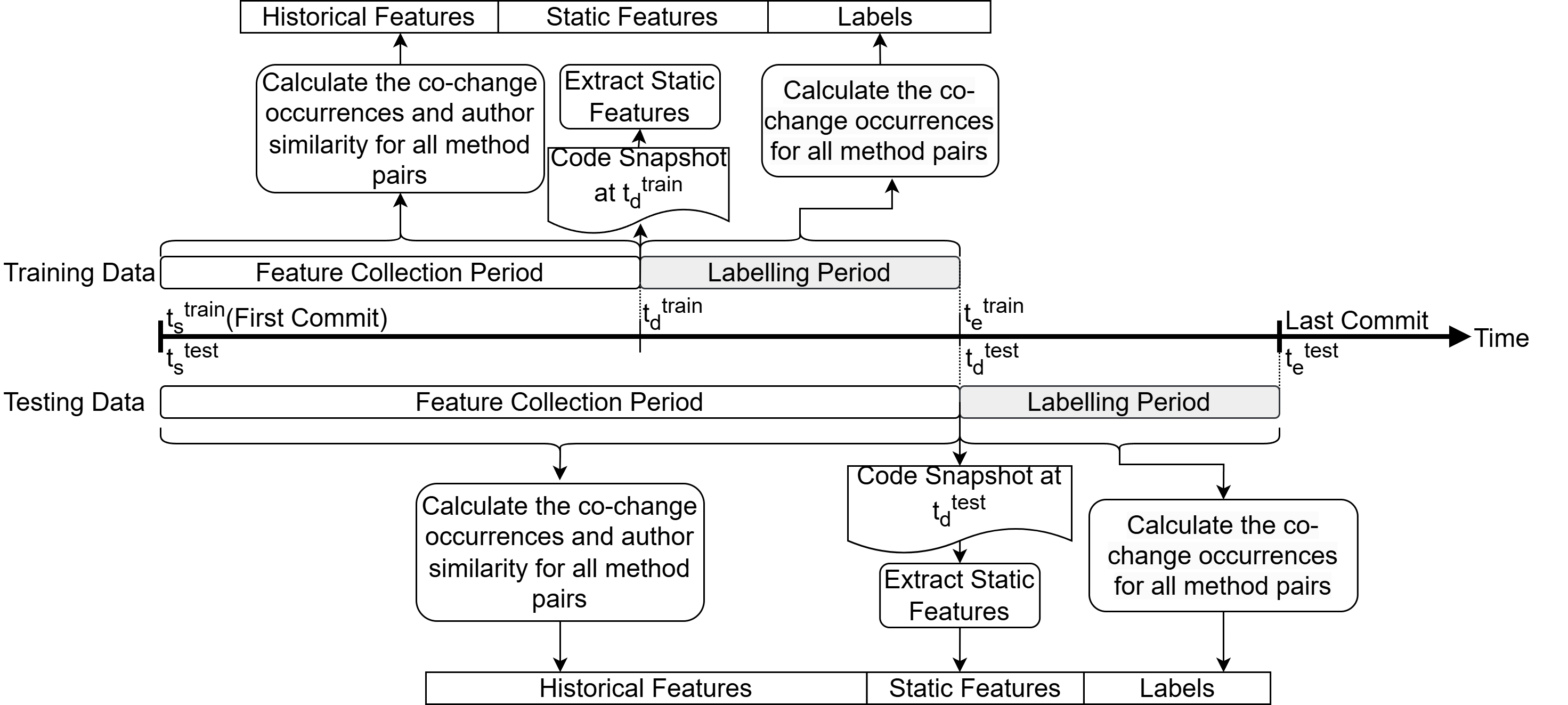}
  \caption{\textbf{Overview of the data creation pipeline for co-change prediction.} This figure illustrates the pipeline for dataset creation as described in Algorithm~\ref{alg:dataset}. The pipeline consists of two main periods: the feature collection period (from \( t_s \) to \( t_d \)) and the labelling period (from \( t_d \) to \( t_e \)). During the feature collection period, historical process and static features are extracted for all method pairs in the repository up to \( t_d \). In the labelling period, co-change occurrences between method pairs are calculated as labels. Temporal separation of feature construction and labeling windows. Feature windows may overlap across training and testing, but labeling windows are strictly disjoint and chronologically ordered to prevent data leakage.}

  \label{img:train-test-periods}
\end{figure}
In real-world scenarios, models are trained with past information and are applied to predict the future behaviour according to the present characteristics. 
As illustrated in Figure~\ref{img:train-test-periods}, for the training phase, predictive features are computed from a 12-month historical window ending at time \( t_d \), followed by a subsequent 6-month window used to construct training labels based on observed co-change occurrences.
For the testing phase, features are computed from a later historical window that occurs entirely before the testing label window, which spans the final 6 months of the repository history.
Importantly, training and testing labeling windows are disjoint and chronologically ordered.

Although the feature collection windows for training and testing may partially overlap, this does not introduce data leakage.
All features are computed exclusively from commits that occurred before their corresponding labeling periods, and no feature incorporates information from its own or any future label window.
Thus, no information from the testing labels is available during training or feature construction.

After the model has been trained, it is tested on a separate set where the predictive features are derived from the historical data from the creation of the repository to 6 months (i.e., 180 days) prior to the final recorded commit. The model tries to predict the candidate methods that are the most likely to be co-changed, and the prediction is compared with the labels based on the co-changes that happened in the last 6 months (i.e., 180 days).

The timeline shown in Figure \ref{img:train-test-periods} indicates that the training and testing datasets are not overlapping and are distinct sets of time-sequential data. This approach is common in time-series forecasting or when dealing with data where temporal dynamics are important~\cite{timeserieseval}, ensuring that the model is evaluated on data it has not seen during training and providing a measure of how well it can generalize to new, unseen data. 
We selected a 12-month training period and a 6-month testing period based on preliminary trial experiments on three projects. This configuration provided a stable and sufficient number of labelled instances for both model training and evaluation.

In our setting, each training or testing instance corresponds to a ranking list constructed for a query method, where candidate methods are ranked based on their likelihood of co-change.
Ranking lists whose candidates all have zero co-change labels are excluded, as they provide no learning information for ranking and are uninformative for evaluation.
As a result, even for large repositories, the number of ranking lists is bounded.
Table~\ref{tab:dataset_stats} reports the descriptive statistics of the ranking lists across all 150 projects for both the training and testing sets.
Both the training and testing datasets are constructed using a 6-month labeling window, which leads to comparable numbers of ranking lists in the two sets.

\begin{table}[tb]
  \caption{Descriptive statistics of ranking lists across all 150 projects.}
  \label{tab:dataset_stats}
  \centering
  \begin{tabular}{*7l}
    \toprule
    Set & Avg. & Min. & 1st Qu. & Median  & 3rd Qu. & Max. \\
    \midrule
    Training & 115.06 & 1 & 15 & 43  & 131 & 1,019 \\
    Testing & 103.11 & 1 & 13 & 35  & 111 & 803 \\
    \bottomrule
  \end{tabular}
\end{table}

To ensure a fair and consistent comparison across models, we tune hyperparameters using a validation set derived exclusively from the training data.
Specifically, we randomly reserve 10\% of the training instances as a validation set and perform grid search to select hyperparameters that maximize NDCG on this validation set.
The testing data is not used at any stage of model selection or parameter tuning.
All models are tuned under the same protocol, ensuring that performance differences are attributable to model capability rather than favorable parameter choices.
The full list of optimized hyperparameters for each model is provided in Section \ref{sec:bg}.

\subsection{Runtime performance and scalability}
The runtime cost of CoRanker can be divided into two stages: feature collection and inference.
Once features are available, inference using the trained learning-to-rank model is lightweight.
In our experiments, generating co-change recommendations typically takes less than one second per query method, even for large repositories containing more than 100K methods.

The primary computational cost lies in feature collection, as historical and structural features must be extracted from version control data.
However, this process can be performed offline and cached.
In practice, only a small subset of features needs to be updated incrementally when a method is modified or when new commits are merged.
As a result, CoRanker is well suited for real-time or near-real-time use in development workflows, such as IDE assistance or pull request analysis.

\begin{algorithm}
\caption{Dataset Creation for Method Co-change Identification}
\label{alg:dataset}
\begin{algorithmic}[1]
\State \textbf{Input:}
\State \quad $t_s$ (repository creation time),
\State \quad $t_d$ (end of feature collection period and the start of labelling period),
\State \quad $t_e$ (end of labelling period)
\State \textbf{Output:} Dataset for co-change prediction
\State $M \gets$ all methods in the repository at time $t_d$ \Comment{$M$: set of all methods}
\State Collect edit history for each method $m \in M$ from $t_s$ to $t_e$ 
\State $M_{\text{valid}} \gets M \setminus$ \{test methods, methods with no edit history\} \Comment{$M_{\text{valid}}$: valid methods with edit history}
\State Initialize dataset $D \gets \emptyset$ \Comment{$D$: co-change identification dataset}

\For{each method $m_q \in M_{\text{valid}}$} \Comment{$m_q$: query method}
    \State Create a ranking list for $m_q$
    \State $M_{\text{candidate}} \gets M_{\text{valid}} \setminus \{m_q\}$ \Comment{$M_{\text{candidate}}$: candidate methods}
    \For{each method $m_c \in M_{\text{candidate}}$} \Comment{$m_c$: a single candidate method}
        \State Add the pair $(m_q, m_c)$ to the ranking list
        \State Calculate historical process features for the pair $(m_q, m_c)$ from $t_s$ to $t_d$
        \State Calculate code features for the pair $(m_q, m_c)$ at time $t_d$
        \State Label the pair $(m_q, m_c)$ with co-change occurrence between $t_d$ and $t_e$
    \EndFor
    \State Add the ranking list for $m_q$ to the dataset $D$

\EndFor

\State Exclude ranking lists where all labels (i.e., co-change occurrence) are $0$  

\Comment{Results will be 0/0 and thus not meaningful}

\end{algorithmic}
\end{algorithm}

\section{Results}
\label{sec:results}
\subsection{RQ1: What is the performance of CoRanker to rank co-changed methods?}
\subsubsection{Motivation}
Identifying the co-change relationships between methods can help the development teams better understand and maintain their systems.
However, existing work identifying the co-changed methods has different limitations (e.g., overlooking the co-change relationships between non-cloned code and identifying co-changes that occurred only in the same commits). 
To address the aforementioned limitations, we propose applying LtR models to identify a small set of methods with the highest probability of being co-changed at the PR level.
In RQ1, we aim to understand the performance of CoRanker in identifying the co-changed methods.

\subsubsection{Approach}
We collect the features from the subject projects as described in Section \ref{sec:expSetup}. 
We train and evaluate the eight models listed in Table \ref{tab:models} using the LtR framework RankLib\footnote{https://sourceforge.net/p/lemur/wiki/RankLib/}.
To evaluate the predictive power of the ranking models, we use the Normalized Discounted Cumulative Gain at position k (NDCG@k) metric~\cite{ndcg} and Precision at position k (precision@k) metric~\cite{liu2009learning,manning2008introduction}. 
NDCG is a well-adapted evaluation metric to measure the overall ranking quality of a model~\cite{ndcg-good}. 
The calculation of NDCG is based on the premise that highly relevant items appearing lower in a search result list should be penalized. 
NDCG is calculated as follows:
$$NDCG@k = \frac{DCG@k}{IDCG@k}$$
where k specifies the number of top positions to consider;
DCG@k (Discounted Cumulative Gain) measures the usefulness of the document based on its position in the result list (lower positions are penalized);
and IDCG@k is the best possible (ideal) DCG given the query.
The calculation of DCG@k is given by the following formula:
$$DCG@k = \Sigma_{i = 1}^{k}{ \frac{2^{rel_i} - 1}{\log_2(i+1)}}$$
where $rel_i$ is the relevance score of the $i^{th}$ item in the ranked list  and $k$ is the number of items in the list.
\begin{comment}
    For example, given a query method $M_q$ and four candidate methods $\{a, b, c, d\}$ that are co-changed with the $M_q$ method $\{2, 1, 1, 0\}$ times, respectively.
The IDCG@3 is 4.13 when the candidate methods are ordered as $\{a, b, c, d\}$ or $\{a, c, b, d\}$, calculated as:
$$IDCG@3 = \frac{2^{2} - 1}{\log_2(1+1)}+\frac{2^{1} - 1}{\log_2(2+1)}+\frac{2^{1} - 1}{\log_2(3+1)} = 4.13$$
Suppose a prediction gives a ranking of $\{b,c,d,a\}$, the DCG@3 is calculated as:
$$DCG@3 = \frac{2^{1} - 1}{\log_2(1+1)}+\frac{2^{1} - 1}{\log_2(2+1)}+\frac{2^{0} - 1}{\log_2(3+1)} = 1.63$$
The NDCG@3 for the prediction is $\frac{DCG@3}{IDCG@3} = 0.40$.

\end{comment}

NDCG@k ranges from 0 to 1, where a score of 1 indicates that the predicted ranking perfectly aligns with the relevance scores, and a score of 0 indicates that no relevant items are present in the top-k results. NDCG@k is widely used in the evaluation of ranking systems because it accounts for both the relevance of items and their position in the ranking \cite{ndcg-good}. 
We chose NDCG for our relevance scores because it effectively handles scalar labels, accommodating varying levels of relevance that other metrics treat as binary. Furthermore, NDCG rewards rankings that place highly relevant items at the top, making it more sensitive to ranking quality than metrics like Top-k accuracy, Recall, or MRR \cite{ndcg}\cite{ndcgtheory}. 
NDCG@k can also be adjusted to different values of k, enabling comparison across varying ranking lengths to determine the most effective configuration. In our analysis, we evaluate the model's performance using k values of 1, 3, 5, and 10.

To comprehensively assess CoRanker's performance, we perform evaluations on each subject project individually by calculating the mean NDCG@k for all query methods. After determining the performance metrics for each project, we compute the mean and median scores over all the projects. 

We also measure ranking performance via \textbf{precision@k}~\cite{liu2009learning,manning2008introduction}, based on a notion of ``top-\(k\) relevance'' from the ground truth. Specifically, for each query method \(M_q\), we define its \textbf{\textit{true co-change relationships}} (\(T@k(M_q)\)) to be those \(k\) candidate methods with the highest co-change occurrences in the ground truth. For example, if a query method \(M_q\) changed 1, 2, 3, 4, 5, 6, 7, 8, 9, and 10 times with 10 candidates, then the five candidates with frequencies 6, 7, 8, 9, and 10 would be labelled as \textbf{\textit{true}} (\(T@k(M_q)\)). 
However, in some cases, there may be fewer than \(k\) such candidates with non-zero co-change frequency. For instance, when \(M_q\) only has three co-change candidates with non-zero co-change counts, then \(T@k(M_q)\) will contain only three elements.

The model’s top-\(k\) predicted candidates are treated as \textbf{\textit{predicted positives \((P@k(M_q))\)}}. The intersection between \((P@k(M_q))\) and \((T@k(M_q))\) yields the \textbf{\textit{true positives}} from the model’s perspective. To account for cases where \(|T@k(M_q)| < k\), we normalize the precision score by the maximum of \(k\) and \(|T@k(M_q)|\). Concretely:
\[
\text{precision@k}(M_q) = 
\frac{|P@k(M_q) \cap T@k(M_q)|}{max(k,|T@k(M_q)|)}.
\]

We then average \(\text{precision@k}(M_q)\) across all query methods \(M_q\) in the test set. This metric captures how many of the \emph{most relevant} co-changed methods (in terms of co-change frequency) the model correctly includes in its own top-\(k\) recommendations.
Although \textit{precision@k}, \textit{recall@k}, and \textit{accuracy@k} are numerically equivalent in this evaluation setting, we adopt the term \textit{precision@k} to reflect the fact that the ground-truth labels are derived based on top-\(k\) relevance ranking.

Furthermore, we want to explore the relationship between the performance and three key metrics of the projects: the number of lines of code (LOC), the number of contributors, and the project's lifespan. For each feature, we divide the projects into 3 groups: 1) the lowest 25\%, 2) median 50\%, and 3) the highest 25\%.
Then, we compare the performance of the model in different groups.

\subsubsection{Results}
\begin{table*}
  \caption{The performance of the studied LtR models using the NDCG@k metric with k equals to 1, 3, 5, and 10.}
  \label{tab:rq1}
    \resizebox{\textwidth}{!}{

  \begin{tabular}{llllllllll}
    \toprule
type       & Model             &   \multicolumn{2}{c}{k=1}&   \multicolumn{2}{c}{k=3}&   \multicolumn{2}{c}{k=5}&   \multicolumn{2}{c}{k=10}\\
       &              & mean  &   median& mean  &   median& mean  &   median& mean  &   median\\
    \midrule
Point-wise &Linear Regression (LR) &	0.7848&	0.8304&	0.8192&	0.8695&	0.8350&	0.8880&	0.8578&	0.9043\\
Point-wise &MART&	0.6794&	0.7868&	0.7446&	0.8111&	0.7751&	0.8668&	0.8024&	0.8750\\
Point-wise &Random Forest (RF) &	\textbf{0.7889}&	\textbf{0.8850}&\textbf{	0.8324}&	\textbf{0.8989}&	\textbf{0.8394}&	\textbf{0.9106}&	\textbf{0.8681}&	\textbf{0.9308}\\
Point-wise &DNN&	0.7110&	0.8099&	0.8057&	0.8613&	0.8052&	0.8601&	0.7947&	0.8549\\
Pair-wise&RankNet&	0.5969&	0.7667&	0.6937&	0.8252&	0.7075&	0.8074&	0.7314&	0.8464\\
Pair-wise  &LambdaMART&	0.7013&	0.7961&	0.7724&	0.8395&	0.7949&	0.8595&	0.8182&	0.8758\\
List-wise  &Coordinate Ascent&	0.7236&	0.8424&	0.7530&	0.8652&	0.7329&	0.8495&	0.7414&	0.8826\\
List-wise  &ListNet&	0.6529&	0.7727&	0.6750&	0.8063&	0.7062&	0.8143&	0.7459&	0.8471\\

  \bottomrule
\end{tabular}
}
\end{table*}

\begin{table*}
  \caption{The performance of the studied LtR models using the precision@k metric with k equals to 1, 3, 5, and 10.}
  \label{tab:rq1prec}
    \resizebox{\textwidth}{!}{

  \begin{tabular}{llllllllll}
    \toprule
type       & Model             &   \multicolumn{2}{c}{k=1}&   \multicolumn{2}{c}{k=3}&   \multicolumn{2}{c}{k=5}&   \multicolumn{2}{c}{k=10}\\
       &              & mean  &   median& mean  &   median& mean  &   median& mean  &   median\\
    \midrule
Point-wise &Linear Regression (LR) &	0.4940&	0.5409&	0.6134&	0.6893&	0.7072&	0.7440&	0.7356&	0.7924\\
Point-wise &MART&	0.4641&	0.7868&	0.5816&	0.7033&	0.6826&	0.7395&	0.6740&	0.7752\\
Point-wise &Random Forest (RF) &	\textbf{0.5274}&	\textbf{0.5793}&\textbf{	0.6811}&	\textbf{0.7292}&	\textbf{0.7418}&	\textbf{0.8036}&	\textbf{0.7822}&	\textbf{0.8301}\\
Point-wise &DNN&	0.4245&	0.6017&	0.5883&	0.6420&	0.6964&	0.7721&	0.7381&	0.8050\\
Pair-wise&RankNet&	0.3528&	0.4420&	0.4792&	0.5259&	0.6583&	0.7446&	0.6740&	0.7434\\
Pair-wise  &LambdaMART&	0.4629&	0.5933&	0.5740&	0.6608&	0.7196&	0.7623&	0.7283&	0.7936\\
List-wise  &Coordinate Ascent&	0.5538&	0.5969&	0.6772&	0.6322&	0.7259&	0.7745&	0.7551&	0.8126\\
List-wise  &ListNet&	0.4356&	0.5721&	0.5179&	0.5933&	0.6021&	0.7180&	0.6959&	0.7429\\

  \bottomrule
\end{tabular}
}
\end{table*}

\begin{figure*}
\centering
  \includegraphics[width=\linewidth]{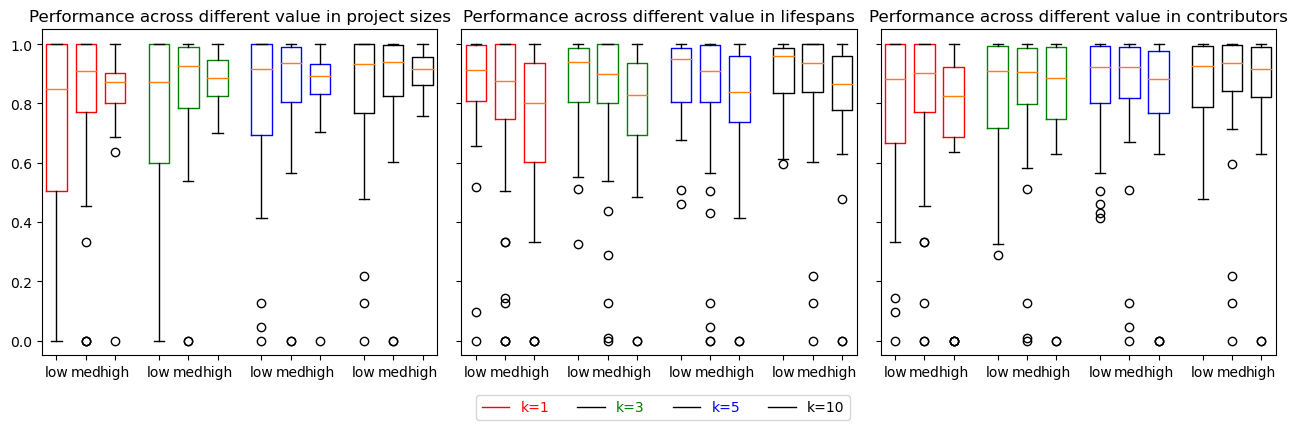}
  \caption{The NDCG@k of the RF model grouped by different characteristics of the studied projects.}
  \label{img:groupedresults}
\end{figure*}

\textbf{The RF model achieves the best performance across both NDCG@k and precision@k metrics.}
Table~\ref{tab:rq1} shows that RF consistently outperforms all other learning-to-rank models across different k values in NDCG@k, achieving the highest mean NDCG@5 score of \textbf{0.8394}. Similarly, in Table~\ref{tab:rq1prec}, RF also achieves the best performance in terms of precision, reaching a \textbf{precision@5 score of 0.7418}. These results confirm the effectiveness of RF in identifying the most relevant co-changed method candidates.

The RF performance indicates its effectiveness in ranking tasks compared to alternatives like LR, MART, RankNet, LambdaMART, AdaRank (which notably failed to converge in most projects), Coordinate Ascent, and ListNet. 
Furthermore, we apply the Wilcoxon signed-rank test~\cite{wilcoxon}, a non-parametric test used to compare paired samples to determine whether their population mean ranks differ. This test is particularly suited to our experiment as it does not assume normality. 
The result confirms the statistical significance of these results, with a p-value less than 0.05. The statistical analysis underscores the robustness of RF in providing more accurate and reliable rankings in comparison to other models evaluated in our study.

Among all the learning-to-rank models evaluated, the RF model achieved the best performance. One possible explanation is that RF is particularly robust to noisy and redundant features, which are likely present in our feature set due to the mix of semantic, historical, and structural code characteristics. RF works by aggregating predictions from multiple independently trained decision trees, which reduces variance and overfitting without relying on sequential learning. This ensemble nature allows it to capture complex, non-linear interactions among features while remaining less sensitive to feature scale or correlation~\cite{liaw2002classification, caruana2006empirical, strobl2009introduction}. These strengths may contribute to RF's consistent performance advantage over models like RankNet, AdaRank, or ListNet, which can be more sensitive to feature quality and hyperparameter settings. While we focus on predictive accuracy in this paper, a deeper investigation into model-specific error patterns and feature contributions will be left for future work.

%Among all the learning-to-rank models evaluated, the Random Forest (RF) model achieved the best performance. One possible explanation is that RF is particularly robust to noisy and redundant features, which are likely present in our feature set due to the mix of semantic, historical, and structural code characteristics. RF works by aggregating predictions from multiple independently trained decision trees, which reduces variance and overfitting without relying on sequential learning. This ensemble nature allows it to capture complex, non-linear interactions among features while remaining less sensitive to feature scale or correlation.
Hence, we use RF as the selected model in the next RQs.

\textbf{The RF model achieves the highest performance of 0.8885 NDCG@5 in identifying the co-changed methods in medium-size and short-term projects.}
Figure~\ref{img:groupedresults} shows the performance of the RF model with respect to different characteristics of the studied projects.
We observe performance improvements in RF models as the value of k increases from 1 to 10.
We also find that the RF model achieves higher performance in identifying the co-changed methods in the medium-size projects compared to small-size and large-size projects. 
The RF performance decreases as the project lifetime increases. 
We conduct the Wilcoxon signed-rank test to compare the RF model performance across different project characteristics. 
The results indicate that, overall, there is no significant difference in model performance among the groups, with the exception that long-term projects exhibit a significantly lower performance compared to young projects. 
This might be attributed to using the entire project history to generate code features, where older historical data could lower the performance of the model.

\subsubsection{Error Analysis}

To better understand the limitations of CoRanker, we conducted a lightweight error analysis focused on the predictions of the RF model. We randomly sampled 97 query methods with incorrect top-5 predictions from the \texttt{netty/netty} repository, using a 95\% confidence level and 10\% margin of error. Manual inspection of these cases revealed three major patterns:

(1) \textbf{Co-changes with many methods:} Some query methods serve core functionality and have historically co-changed with many other methods. In these cases, CoRanker may recommend a method that is indeed co-changed, but not among the top-5 most frequent ones. This highlights a limitation of binary evaluation metrics such as precision@k and reinforces the value of NDCG@k, which captures ranking quality in scalar-labelled data.

(2) \textbf{Outdated co-change features:} In several cases, a query method previously had strong co-change ties with a candidate method during the early phases of the project, but these relationships no longer exist. However, CoRanker still prioritized them due to their historical occurrence. This observation is consistent with Figure~\ref{img:groupedresults}, where the performance drop in projects with longer lifespans, suggesting that future work could incorporate recency-aware features, such as the number of days since last co-change.

(3) \textbf{Sparse co-change history:} In some cases, the query method had never co-changed with other methods in the past and only appeared 0–1 times in co-changes during the labelling period. This lack of historical signals limits the model's ability to make accurate predictions. Introducing a confidence-based filtering mechanism to suppress such low-signal cases may improve the model’s robustness when applied in real-world scenarios.

These findings offer actionable insights for improving CoRanker in future iterations, such as temporal decay of historical features and filtering heuristics for low-signal predictions.

\begin{boxA}    
The RF model consistently outperforms other LtR models with the highest average scores: \textbf{NDCG@5 of 0.8394} and \textbf{precision@5 of 0.7418}. 
The RF model achieves the highest performance in identifying the co-changed methods in medium-size and short-term projects.
\end{boxA}

\subsection{RQ2: Can CoRanker perform better than baselines? }
\subsubsection{Motivation}
Prior work uses code-clone and association rules techniques to rank the co-changed methods on the commit level. 
To our knowledge, our study is the first attempt to apply ML technique (i.e., LtR techniques) for ordering co-change methods at a PR level. 
This results in a lack of existing benchmarks for comparison.
In this RQ, we compare CoRanker with six baselines.
By comparing the results of CoRanker to existing baselines, we aim to determine whether CoRanker is a significant improvement over the existing methods. 

\subsubsection{Approach}

Given a changed method $M_q$ and a group of candidate methods $M_C$, the six baselines are defined as follows:

\noindent \textbf{1) Support ranking:} orders the co-change candidates $M_C$ of a query method $M_q$ by the number of times each candidate has co-changed with $M_q$ in the past. This baseline represents a heuristic, frequency-based ranking approach derived from association rules (e.g., HistoRank~\cite{historank}), where the importance of historical co-change is fixed and not learned from data.

\noindent \textbf{2) File proximity ranking:} ranks the candidates $M_C$ by their file path proximity to the query method $M_q$. 
The file path proximity is the distance between the file entities in the file system tree. Files with the same file proximity are ranked by their support ranking. The baseline is based on association rules from HistoRank ~\cite{historank}. 
    
\noindent \textbf{3) Using a code clone tool:} This approach uses NiCad clone-detector ~\cite{nicad} as a co-change detector used by Mondal et al. ~\cite{EConClone2,EConClone3}. It ranks the candidate methods $M_C$  by the code similarity score given by the NiCad clone detector~\cite{nicad}.

\noindent \textbf{4) FCP2Vec:} This approach uses Word2Vec to encode file paths into vectors and applies k-NN to retrieve the top k nearest candidates with the highest probability of co-changes\cite{fcp2vec}. To adapt it for method-level prediction used in our work, we append the method name to the file name.

\begin{figure}[tb]
    \centering
    \includegraphics[width=\linewidth]{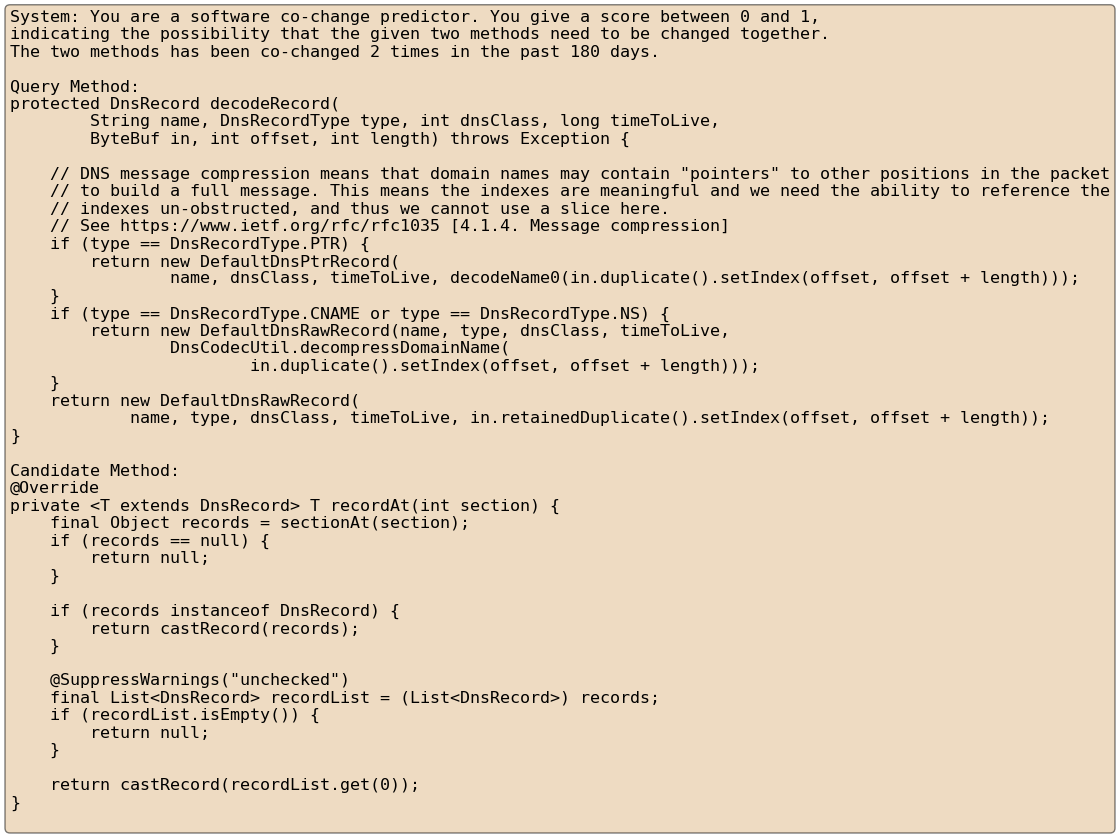}
    \includegraphics[width=\linewidth]{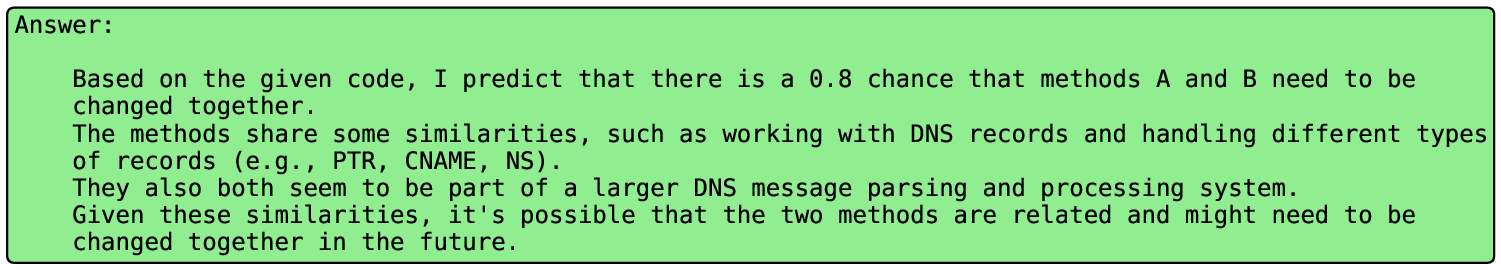}
    \caption{Example prompt used by the co-change predictor. The historical co-change information (e.g., “The two methods has been co-changed 2 times in the past 180 days.”) is included only in the history-aware setting and is omitted in the zero-shot setting.}
    \label{fig:prompt}
\end{figure}

\noindent \textbf{5) StarCoder 2 - Scalar (Zero-shot / History-aware):} We implement a point-wise LtR model that utilizes the open-source LLM, StarCoder 2, with 15B parameters and 16-bit precision\cite{starcoder2}. StarCoder 2 is particularly well-suited for this task as it has been pre-trained on a wide range of programming languages, enabling it to understand both the syntax and deeper structural patterns in code\cite{starcoder2}. The model takes the source code of the query method $M_q$ and candidate method $M_c$ as input and predicts a float between 0 and 1, representing the probability of co-change. An example prompt and the corresponding predictions are listed in Fig. \ref{fig:prompt}. Candidates are then ranked based on their predicted probabilities. Due to limitations in computing resources, we applied this baseline to 45 randomly selected projects from our dataset, comprising over 2.4 million method pairs in total.

We evaluate this baseline under two information settings.
In the \emph{zero-shot} setting, the prompt contains only the source code of $M_q$ and $M_c$, without any explicit historical co-change information.
In the \emph{history-aware} setting, we augment the prompt with explicit historical co-change statistics (e.g., the number of past co-change occurrences between $M_q$ and $M_c$), allowing us to assess whether StarCoder 2 can effectively leverage such signals when they are made available.

\noindent \textbf{6) StarCoder 2 - Binary (Zero-shot / History-aware):} 
Prior work~\cite{logitsRank,qian2025tokenpromptrobustzeroshotclassification,lepagnol2024smalllanguagemodelsgood} demonstrated that Large Language Models (LLMs) can effectively perform zero- or few-shot classification by comparing the log-probabilities of different class tokens in a prompt. Building on that approach, we investigate the feasibility of employing a similar methodology in this work. This baseline also leverages the open-source LLM, StarCoder 2 \cite{starcoder2}, but formulates co-change prediction as a binary (\textit{yes} / \textit{no}) classification. For each pair \((M_q, M_c)\), we prompt StarCoder 2 with a direct yes/no question asking whether \(M_c\) has a co-change relationship with \(M_q\).
We then extract the probabilities of StarCoder 2's first generated token for both ``yes''/``Yes'' and ``no''/``No.'' Concretely, we define:
\[
P(\text{yes}|(M_q,M_c)) = \frac{P(\texttt{token == "yes"}|(M_q,M_c)) + P(\texttt{token == "Yes"}|(M_q,M_c))}{2},\]
\[
P(\text{no}|(M_q,M_c)) = \frac{P(\texttt{token == "no"}|(M_q,M_c)) + P(\texttt{token == "No"}|(M_q,M_c))}{2}.
\]
Next, we compute the following ratio-based score:
\[
\text{score} = \frac{P(\text{yes}|(M_q,M_c)) + 1}{P(\text{no}|(M_q,M_c)) + 1},
\]
and rank the candidate methods based on this score in descending order. Similar to baseline 5 StarCoder 2 - Scalar, we evaluate this baseline on the same 45 randomly selected projects, leading to predictions for over 2.4 million method pairs in total.

As with the scalar formulation, we evaluate this baseline under two information settings.
In the \emph{zero-shot} setting, the prompt includes only the source code of $M_q$ and $M_c$.
In the \emph{history-aware} setting, we additionally provide explicit historical co-change statistics within the prompt.

In the baselines, the co-change methods are identified at the commit and file levels, whereas our work is based on PR levels. We adopt the baselines and re-implement them at the PR level.

We selected StarCoder 2 due to its strong performance among open-source code models and its accessibility for large-scale evaluation across real-world projects~\cite{starcoder2, chen2023huggingface}. Although more recent LLMs such as GPT-4 may offer higher performance on some tasks, they are either not open-source or not feasible to run across millions of code pairs due to their substantial computational requirements. Our goal is to explore the realistic applicability of LLMs in a scalable, reproducible setting, and StarCoder 2 strikes a balance between model capability and practical deployment.

\subsubsection{Results}
\textbf{CoRanker shows a significant improvement in predicting the co-changed methods over six established baselines by a margin of 4.7\%--573.5\% in terms of NDCG@5.}
The comparison between the baselines is shown in Table~\ref{tab:baselines}. 
We find that CoRanker outperforms the baseline approaches, with a 10.2\% difference in NDCG@1, a 7.9\% difference in NDCG@3, a 4.7\% difference in NDCG@5, and a 4.3\% difference in NDCG@10 compared to the best-performing baseline, support ranking.
We also conduct the Wilcoxon signed-rank test to determine the significance of the differences between CoRanker and the baseline approaches in performance. 
The results show that CoRanker is significantly better than the baselines.

We observe that the differences between CoRanker and the baseline approaches increase as the value of k decreases. 
This result indicates that CoRanker is more powerful than baselines in making precise predictions for the small number of the selected candidate methods k.

\textbf{The support ranking approach achieves the best performance in the baseline approaches.} 
The support ranking reaches a relatively good performance as the support ranking has the same definition as the labels (i.e., the support ranking and the labels measure the number of co-changes between the query method $M_q$ and the candidate method $M_c$ in the training and testing periods respectively). 
File proximity ranking and clone ranking have a limited ability to detect co-changes, achieving NDCG@5 scores of 0.64 and 0.12, respectively.

Although the average improvement over support-based ranking is moderate, the gains are substantially larger at early ranks, with improvements of approximately 10.2\% at $k=1$ and 7.9\% at $k=3$.
These early-rank improvements are particularly important in practice, as developers typically inspect only a small number of top-ranked recommendations.
This indicates that the learning-to-rank framework is especially effective at refining the ordering among top candidates with comparable historical co-change frequency—precisely the cases where heuristic ranking provides limited discrimination.

The clone detector has limited ability to accurately predict co-changes because the number of detected clone pairs is very small compared to the total number of candidate entities.

We further observe clear performance differences among the LLM-based baselines under different output formulations and information settings.
In particular, providing historical co-change information improves the performance of the scalar formulation of StarCoder 2 compared to its zero-shot counterpart, whereas the history-aware binary formulation performs worse than the zero-shot binary setting.
Moreover, even the best-performing LLM configuration remains consistently below the heuristic support-ranking baseline.

\textbf{Exploration of LLM-based co-change prediction.}
Although large language models such as StarCoder 2 have demonstrated strong capabilities in tasks like code generation and summarization, their performance on co-change ranking is consistently lower than that of feature-based approaches.
As shown in Table~\ref{tab:baselines}, augmenting StarCoder 2 with historical co-change information improves the scalar formulation compared to its zero-shot variant.
However, even in this history-aware setting, the scalar LLM baseline remains inferior to the simple support-ranking heuristic, indicating that the model is unable to fully exploit historical features for accurate prioritization.

In contrast, the binary formulation of StarCoder 2 exhibits degraded performance when historical information is introduced.
One possible explanation is that explicitly injecting historical statistics into a binary yes/no prompt can overly bias the model’s token-level decision, leading to unstable probability estimates and reduced ranking quality.
This suggests that LLM-based binary classification may be particularly sensitive to prompt design when applied to fine-grained evolutionary prediction tasks.

Overall, these results indicate that while LLMs can benefit from explicit contextual information, they are not well-suited for modeling co-change prediction as a ranking problem.
Unlike feature-based learning-to-rank models, LLMs are not designed to directly capture evolutionary coupling patterns from historical version control data.
Consequently, their performance remains limited even when such information is explicitly provided.

In addition to lower predictive performance, LLM-based approaches incur substantially higher computational costs.
In our experiments, evaluating StarCoder 2 on 45 projects involving over 2.4 million method pairs required approximately 2.5 seconds per pair on an NVIDIA A100 GPU with 80GB memory.
Such computational overhead makes LLM-based co-change prediction impractical for large-scale or time-sensitive development scenarios.
In contrast, CoRanker relies on lightweight feature extraction and efficient learning-to-rank models, enabling scalable deployment in real-world software maintenance workflows.

\begin{boxA}    
    CoRanker shows a significant improvement in performance over six established baselines by a margin of 4.7\% --573.5\% in terms of NDCG@5, particularly when the number of recommendations (i.e., k) is smaller. Notably, the StarCoder 2 baseline lags behind the RF model, implying that large code-focused language models trained without explicit co-change knowledge may not capture evolutionary coupling patterns effectively.
\end{boxA}

\begin{table}[tb]
  \caption{Comparison of CoRanker with baselines.}
  \label{tab:baselines}
  \centering
  \begin{tabular}{l|l|l|l|l|l}
    \toprule
    k & RF Model & Baseline       & Mean   & Difference  & Significance \\
    \midrule
    \multirow{8}{*}{1}  & \multirow{8}{*}{0.7898} & Support      & 0.7166 & 10.2\%  & *  \\
                        &                        & File Proximity & 0.5487 & 43.9\%  & ** \\
                        &                        & Code Clone   & 0.0014 & 57528.6\% & ** \\
                        &                        & FCP2Vec      & 0.4025 & 96.2\%  & ** \\
                        &                        & StarCoder 2 - Scalar - zero-shot& 0.5294 & 49.2\%  & ** \\
                        &                        & StarCoder 2 - Scalar - history-aware& 0.6024 & 31.1\%  & ** \\
                        &                        & StarCoder 2 - Binary - zero-shot & 0.4613 & 71.2\%  & ** \\
                        &                        & StarCoder 2 - Binary - history-aware & 0.2993 & 163.9\%  & ** \\
    \midrule
    \multirow{8}{*}{3}  & \multirow{8}{*}{0.8330} & Support      & 0.7720 & 7.9\%   & *  \\
                        &                        & File Proximity & 0.6073 & 37.2\%  & ** \\
                        &                        & Code Clone   & 0.0847 & 883.8\% & ** \\
                        &                        & FCP2Vec      & 0.4483 & 85.8\%  & ** \\
                        &                        & StarCoder 2 - Scalar - zero-shot & 0.5630 & 48.0\%  & ** \\
                        &                        & StarCoder 2 - Scalar - history-aware & 0.6269 & 32.9\%  & ** \\
                        &                        & StarCoder 2 - Binary - zero-shot & 0.4837 & 72.2\%  & ** \\
                        &                        & StarCoder 2 - Binary - history-aware & 0.3102 & 168.5\%  & ** \\
    \midrule
    \multirow{8}{*}{5}  & \multirow{8}{*}{0.8399} & Support      & 0.8023 & 4.7\%   & *  \\
                        &                        & File Proximity & 0.6364 & 32.0\%  & ** \\
                        &                        & Code Clone   & 0.1247 & 573.5\% & ** \\
                        &                        & FCP2Vec      & 0.4620 & 81.8\%  & ** \\
                        &                        & StarCoder 2 - Scalar - zero-shot & 0.5951 & 41.1\%  & ** \\
                        &                        & StarCoder 2 - Scalar - history-aware & 0.6727 & 24.9\%  & ** \\
                        &                        & StarCoder 2 - Binary - zero-shot & 0.5277 & 59.2\%  & ** \\
                        &                        & StarCoder 2 - Binary - history-aware & 0.3580 & 134.6\%  & ** \\
    \midrule
    \multirow{8}{*}{10} & \multirow{8}{*}{0.8684} & Support      & 0.8328 & 4.3\%   & *  \\
                        &                        & File Proximity & 0.6729 & 29.1\%  & ** \\
                        &                        & Code Clone   & 0.1958 & 343.6\% & ** \\
                        &                        & FCP2Vec      & 0.5264 & 65.0\%  & ** \\
                        &                        & StarCoder 2 - Scalar - zero-shot  & 0.6004 & 44.6\%  & ** \\
                        &                        & StarCoder 2 - Scalar - history-aware  & 0.7104 & 22.2\%  & ** \\
                        &                        & StarCoder 2 - Binary - zero-shot & 0.5760 & 50.8\%  & ** \\
                        &                        & StarCoder 2 - Binary - history-aware & 0.4115 & 111.0\%  & ** \\
\bottomrule
\multicolumn{6}{l}{*p<0.05\,\, **p<0.005}
\end{tabular}
\end{table}

\subsection{RQ3: What are the important features for building the LtR models?}

\subsubsection{Motivation}
It is important to understand the features (e.g., path similarity and author similarity) that would influence the predictive ranking the most.
Such insights improve the interpretability of CoRanker, which helps development teams better understand and rely on our predictions.
In addition, understanding  the most important features can assist practitioners in better prioritizing their efforts by focusing on collecting and analyzing such features. 
Hence, in this RQ, we analyze the importance of the studied features in building the LtR models.

\subsubsection{Approach}
Permutation importance is a technique used to evaluate the importance of each feature in a machine-learning model.
The basic idea behind permutation importance is to randomly shuffle the values of a feature, and then observe the decrease in the performance of the model. 
The larger the decrease in performance, the more important the feature is considered to be~\cite{randomforest}.

We calculate the permutation importance of each of the collected features and compare their importance. 
The permutation importance is calculated by training a model on the original dataset.
Then, shuffling the values of a specific feature and re-evaluating the performance of the model on the shuffled data. 
The difference in performance between the model trained on the original data and the model trained on the shuffled data is the permutation importance of that feature.
For simplicity, we limit our evaluation of feature importance to the setting where k = 5.
We choose k=5 as it provides a balanced choice among the experimented K values (i.e., 1,3,5 and 10) with a reasonable number for the recommendation of co-changes.

\subsubsection{Results}

\begin{table}[tb]
  \caption{The permutation importance of the studied features.}
  \label{tab:rq3}
  \begin{tabular}{ll}
    \toprule

Feature & Importance \\
    \midrule
Number of co-changes        & 0.3844                                  \\
Path similarity             & 0.0539                                   \\
Authors similarity          & 0.0392                                    \\
Code dependency           & 0.0082                                    \\
Hierarchy similarity      & 0.0043                                   \\
Clone similarity          & 0.0010                                       \\
Argument type similarity   & 0.0010                                   \\
Semantic similarity       & 0.0008                                \\
Argument name similarity       & -0.0007            \\

\bottomrule
\end{tabular}
\end{table}
\textbf{The number of co-changes is the most important feature, with feature importance 0.38.
Path similarity and author similarity also positively influence the model’s performance, with feature importance of 0.05 and 0.04, respectively.}
Table \ref{tab:rq3} shows the importance of the studied features. 
We find that the number of co-changes feature is the dominant one among all the studied features. 
The model trained on the dataset where the number of co-changes is shuffled reached only 0.4692 NDCG@5, showing that the feature has 0.3844 permutation importance.
The rest of the features are not as important as the number of co-changes.
However, we observe that the studied features still contribute positively to the prediction power of the model. 
We also notice that path similarity and authors similarity features are also important to the model.
However, we find that the argument name similarity feature has a negative impact on the prediction power of the model. 
A possible explanation is that argument names are often chosen in a highly local or context-dependent manner and may follow project- or developer-specific naming conventions.
As a result, high argument name similarity does not necessarily imply a functional or evolutionary relationship between methods, and may instead introduce noise when combined with stronger historical features.
In addition, argument names tend to be short and generic (e.g., \texttt{data}, \texttt{value}, \texttt{result}), which further limits their discriminative power.
The slightly negative permutation importance suggests that this feature provides little useful information and may even interfere with the ranking learned from more informative features.
From a practical perspective, this finding indicates that argument name similarity can be safely omitted without degrading model performance.

\textbf{Code similarity features (i.e., code clone and semantic similarity) have low importance in predicting the co-changed methods.}
We find that the importance of clone similarity is 0.001 which is lower than our expectation. 
Although clone detection is found to be a good indicator of co-change relationships as in prior work\cite{EConClone4,EConClone2}, the number of positively detected clones is too small compared to the total amount of methods in our subject systems. 
Hence, the clone similarity does not play an important role in detecting the co-change relationships.
The impact of the semantic similarity of the method is also low (i.e., 0.0008). 
It is possible that the CodeBERT model requires fine-tuning to be adapted to CoRanker. 
Another possibility is that measuring the cosine similarity between the vector embeddings of the query method $M_q$ and the candidate method $M_c$ does not accurately identify the semantic similarity of the methods.

\begin{boxA}
    The number of co-changes emerges as the most important feature, with path similarity and authors similarity also positively influencing the model’s predictive ability. Other features have limited impact, with the argument name similarity feature negatively affecting predictions.
\end{boxA}

\subsection{RQ4: How soon can the model reach a consistent and accurate performance?}

\subsubsection{Motivation}

In the previous RQs, we used 180 days (i.e., 6 months) of edit history to create the training labels and 180 days to create the testing labels, as shown in Figure \ref{img:train-test-periods}. 
However, the selection of 180 days may not be the best choice for all projects. 
The choice of the periods to label the data can be influenced by the projects' updating pace and lifespan. 
More specifically, if the training data is extracted from a long period of history, the model might be trained on obsolete data, leading to inaccurate predictions. 
On the other hand, training the LtR models with a short period of history may not provide enough characteristics about the nature of the co-changed methods which leads to inaccurate predictions. 
In this RQ, we aim to understand how the recency of the training data and period of the training and testing data affect the  performance of the model.  
Therefore, we want to suggest to the practitioners the most suitable period for using the model before it comes necessary to retrain the model with more up-to-date data to maintain the high performance of the model.

\begin{figure}[tb]
  \centering
  \includegraphics[]{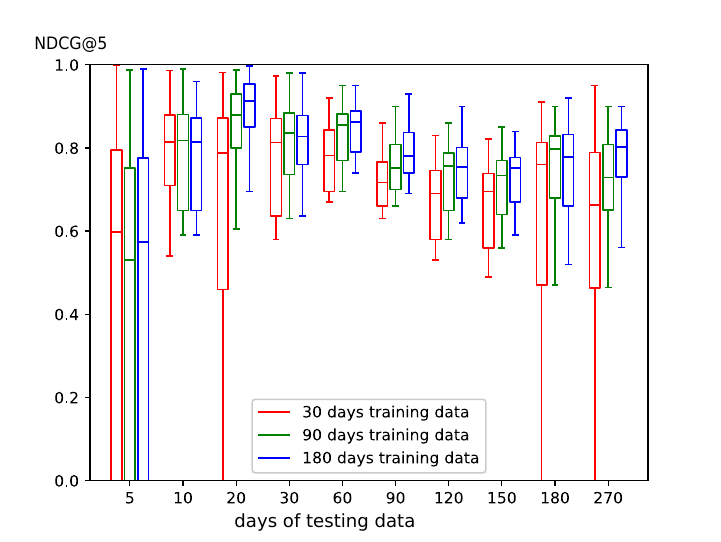}
  \caption{The performance of the RF model with different periods for creating the training and testing labels. 
  }
  \label{img:rq4}
\end{figure}

\subsubsection{Approach}
In our experiment for this RQ, we alter the periods to label the training and testing data to test the performance of the RF model (i.e., the best performing model) with different setups of periods.
Given our time constraints, we focus on a random sample of 30 projects for our analysis in this RQ.

Based on our preliminary analysis, very short labeling periods (e.g., 5–20 days) often contain insufficient co-change information to construct meaningful training data. We therefore use 30 days as the minimum labeling period for training data construction. The training labeling periods of 30, 90, and 180 days are chosen to represent scenarios where development teams train models using monthly, quarterly, and semi-annual historical data, respectively.
For each project, we try different combinations of periods as follows, resulting in a total of 30 different setups:
\begin{itemize}
    \item Period to label the training data: 30, 90, 180 days
    \item Period to label the testing data: 5, 10, 20, 30, 60, 90, 120, 150, 180, 270
\end{itemize}

We use the Wilcoxon signed-rank test with a significance level of 0.05 to determine whether the performance of the model in different settings differs significantly.

\subsubsection{Results}
\textbf{The performance of the RF model begins to decline after 60 days, suggesting a need to retrain models bi-monthly (i.e., 60 days) for optimal results.}
Figure~\ref{img:rq4} shows the performance of the RF model with different periods for creating the training and testing labels.
We find that the performance of the model on five days of testing data is unstable because there is not enough data in the testing dataset to provide an accurate evaluation of the performance.
The models with different training periods reach a relatively high performance after there are enough testing data, and the high performance of the models starts to drop after the testing period reaches 60 days. 
After 60 days, the models lose their prediction power gradually.

\textbf{Models trained over longer periods (90 and 180 days) exhibit higher and more stable performance compared to the models trained with a 30-day training period. }
We find that the models with 90-day and 180-day training periods perform better than the model with a 30-day training period, with no statistically significant difference in predictive power between the 90-day and 180-day models. When evaluated with 270 days of testing data, the model with 180-day training data outperforms the model with 30-day training data by a margin of 20.7\%, and this difference is statistically significant with a p-value less than 0.05. Moreover, the model with a 30-day training period is less stable, as it fails to generate any training data in some projects, resulting in a performance of zero. This instability makes the model with a 30-day training period unreliable, as it cannot be used on certain projects. In our experiments, this issue primarily occurred with the model using a 30-day training period.

\begin{comment}

We find that the models with 90 days and 180 days training period have a better ability to rank the co-change candidates compared to the model with 30 days training period. This difference is statistically significant when evaluated with 20 days and 270 days of testing data, with a p-value less than 0.05.
When evaluated with 270 days testing data, the model with 180 days of training data outperforms the other two models by a margin of 9.8\% and 20.7\%, indicating that the model is less prone to performance degradation.
%We also find that the performance of the model with 180 days of training data has a more consistent and reliable performance than the others (i.e., the model with 180 days of training data drops in a smaller amount than the other two models as the testing period increases). 
However, as depicted in Figure \ref{img:rq4}, the performance of the model, when trained on 30 days of data, exhibits greater instability compared to the performance observed with training periods of 90 and 180 days.
We also observe that, on some projects, no training data is generated when the model uses a 30-day training period, resulting in the model's performance being zero.
In these cases, the model cannot be used on the project. In our experiment, this problem happens mostly on the model that is using 30 days to label the training data. 
\end{comment}
\begin{boxA}
    Models trained over longer periods (90 and 180 days) exhibit higher, more stable performance compared to a 30-day training period. 
    The performance of the model begins to decline after 60 days of the testing period, suggesting a need to retrain models bi-monthly for optimal results.
\end{boxA}

\section{Implications}
\label{sec:implication}

\begin{figure}[tb]
  \centering
  \includegraphics[width = \linewidth]{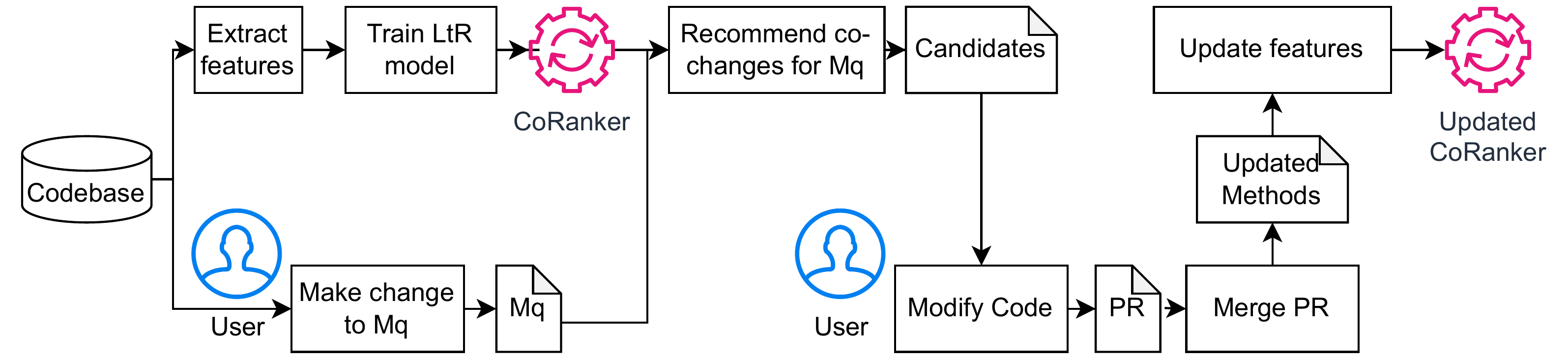}
  \caption{Illustration of the CoRanker workflow for recommending co-changed methods during software development.
  }
  \label{img:implication}
\end{figure}
\subsection{Implications for developers}

Beyond predictive accuracy, CoRanker is designed to be easily integrated into modern software development workflows.
In the following, we describe concrete usage scenarios illustrating how CoRanker can assist developers during code editing, pull request creation, and continuous integration.

\textbf{Developers can use CoRanker to avoid bugs and errors during development.}

There are two practical integration modes in practical development environments:
\textbf{(1) IDE-time assistance} and \textbf{(2) background analysis during pull requests or CI pipelines}.

CoRanker can increase the developers' awareness of co-changed methods and recommend the methods for developers to inspect.
If two methods tend to change together frequently, changing one method without considering the impact on the other method could introduce bugs and errors. By repeatedly identifying method pairs that are co-changed yet live in distant packages or modules, teams gain clues about potential structural “tangles” or architecture drift, thereby improving modularity and maintainability.
There are two ways CoRanker can be used in practical development environments. First, it can be built into an IDE like Eclipse. Developers can download and run the trained model locally, and the IDE can show real-time suggestions as they write or review code. 
Figure~\ref{img:implication} illustrates the intended workflow for integrating CoRanker into practical development environments. A development team first decides to adopt CoRanker and spends approximately 1–2 hours creating the dataset and training the initial learning-to-rank model. During daily development, when a developer modifies a query method \(M_q\) in the IDE, CoRanker uses the trained model to recommend 3–5 likely co-changed candidate methods within about 1 seconds based on the measurement from our study, helping the developer ensure that no relevant changes are overlooked. After the developer completes the modification and merges the pull request, the server automatically computes incremental updates to historical and structural features using minimal computational resources. To reflect evolving co-change patterns, the model is retrained periodically, typically every 2–3 months, as derived from the findings in RQ4, or other intervals depending on the amount of new pull requests submitted. Although the initial data collection process is time-consuming, it is performed only once; afterward, CoRanker makes predictions quickly, and the dataset is updated using only incremental changes, making the approach suitable for practical deployment. This workflow demonstrates that CoRanker can be integrated into IDE-based assistance and pull request or CI pipelines with low overhead and without disrupting existing development practices.

Second, teams can run the model as a background process. For example, suppose a team modifies a set of methods during the day, denoted as \( MC = \{M_1, M_2, \dots, M_n\} \). For each method \( M_i \in MC \), CoRanker predicts a set of likely co-changed methods \( CI(M_i) \). The system then checks whether the predicted set \( CI(M_i) \) is fully included in the actual modified set \( MC \). If any predicted co-changed methods are missing, it issues a warning. For instance, if the team modifies methods \( M_1 \) and \( M_2 \), and the model predicts \( CI(M_1) = \{M_1, M_3, M_4\} \) and \( CI(M_2) = \{M_1, M_3, M_5\} \), the system will detect that \( M_3, M_4, \) and \( M_5 \) were not modified. It will then generate warnings such as: for \( M_1 \), ``you may have missed co-changes to \( M_3 \) and \( M_4 \),'' and for \( M_2 \), ``you may have missed co-changes to \( M_3 \) and \( M_5 \).'' These alerts help developers catch overlooked changes that could lead to defects, particularly in large systems where related entities are distributed across different modules. This background analysis can be integrated into continuous integration (CI) pipelines or pull request checks, where warnings are surfaced before code is merged.
The stable performance across different training windows observed in RQ4 suggests that periodic retraining (e.g., every 2–3 months) is sufficient in practice.

\textbf{Developers can use CoRanker to improve the quality of code.}
Co-changed methods are not necessarily statically coupled—they may belong to different packages, have no explicit dependencies, and still frequently evolve together due to hidden or logical relationships. As a result, identifying them manually can be difficult. By applying CoRanker, developers can uncover these implicit relationships and better understand why certain methods often change together. This insight can guide several possible actions: refactoring the code to reduce hidden dependencies, grouping the co-changing methods into the same file or package to improve modularity, or documenting the relationship to ensure they are updated consistently in the future. 
For example, during major pull requests, feature integrations, or release preparations, the development team can run CoRanker to scan the codebase and detect method pairs with frequent co-changes. This check helps the team identify structural risks and make informed decisions about whether to refactor, reorganize, or track the relationships for safer ongoing maintenance.

\textbf{CoRanker enhances developers' understanding of code relationships.} By identifying co-changed methods at PR, CoRanker enhances the understanding of code relationships, helps developers to better plan and implement changes (e.g., assign the co-changed methods to the same developers), reducing the risk of overlooking necessary modifications that might lead to defects or unstable code at the commit-level.

\subsection{Implications for code reviewers and testers}

\textbf{Code reviewers can use CoRanker as a tool to identify potential side-effects or dependencies between methods.} 
If two methods are frequently co-changed, it indicates that they have a high degree of coupling, and changes to one method may have unintended consequences on the other method.  For example, CoRanker (using K=3) can provide an automated prompt that, upon opening a pull request that changes only method A, “Methods X, Y, Z are typically co-changed with the code you just modified—please check or modify them.” It can work like a real-time suggestion right inside the code review.

\textbf{Adapting for automated test generation or test selection}  While the paper focuses on method co-changes, a direct extension is to identify which tests might be relevant when a certain method changes. If one method is changed, it is more likely that the methods that have a co-change relationship with the changed method can be impacted. Automated test frameworks could highlight or run only those test methods with strong co-change ties to the edited code, thus speeding up continuous integration pipelines.

Together, these scenarios demonstrate that CoRanker can be deployed as a lightweight assistant across IDEs, pull request workflows, and CI pipelines, without requiring intrusive changes to existing development practices.

The choice of K depends on the usage context and the tradeoff between noise and coverage. When the tool is integrated into IDEs, a smaller number of recommended candidates can reduce distraction for developers; in such interactive settings, K values around 3–5 are a reasonable choice. During code review, especially in large repositories, missing a relevant co-change may be more costly and entities may have more valid co-change relationships. In these scenarios, using a larger K can improve coverage and may be preferred.

\subsection{Co-change Beyond Direct Dependencies}
To better understand the types of relationships captured by CoRanker, we conducted a small-scale manual analysis. We randomly selected four repositories and, from each repository, randomly sampled five ranking lists, resulting in 20 ranking lists in total. For each ranking list, we inspected the top-ranked co-changed method pair. We then categorized these pairs into four groups: (1) methods within the same class, (2) methods with direct call dependencies, (3) methods without direct dependencies but sharing similar functionality, and (4) methods with no obvious direct dependency.

Among the 20 inspected pairs, 15 exhibit direct dependencies, including 12 pairs within the same class and 3 pairs connected by direct method calls. The remaining five pairs do not have explicit structural dependencies. Notably, these pairs are primarily related to memory management, multithreading, and cross-package communication (e.g., handlers). Although they lack direct dependencies, they implement closely related functionality and therefore need to be modified together in practice.

These observations suggest that CoRanker can identify co-change relationships beyond explicit structural dependencies. It can reveal co-change relationships with implicit or functional connections that are difficult to detect using static analysis alone. As a result, it can help developers better understand the functional structure of a repository and alert them when relevant co-changed entities are missed.

\section{Threats to Validity}
\label{sec:threat}
In evaluating CoRanker across 150 open-source projects, we recognize several potential threats to the validity of our findings. These threats are categorized into three main types: construct validity threats, external validity threats and internal validity threats.

\textbf{Construct Threats: }In this work, a co-change relationship is identified based on repeated co-change occurrences between methods in pull requests.
This definition captures co-change relationships that are reflected in actual development activities.
However, it is possible that some methods have a co-change relationship but are rarely modified during the observed period, particularly in stable or mature parts of a repository, and therefore do not appear in the version history.
We mitigate this threat by focusing on repositories with active development and by defining co-change relationships based on recurring co-change occurrences rather than single events.

In addition, a construct validity threat arises from evaluating co-change relationships using ranking-based metrics (i.e., NDCG@k).
While a co-change relationship is inherently relational, our evaluation measures the effectiveness of prioritizing co-changed methods in ranked recommendation lists.
This formulation directly reflects the practical goal of recommending relevant co-changed methods to developers.
Since developers typically inspect only a small number of top-ranked candidates, ranking effectiveness provides a meaningful and appropriate operationalization for evaluating co-change detection and ranking performance.

\textbf{External Threats: }Although 150 projects are generally more than the existing studies, it still can be insufficient to generalize our experiment results to all the Java software projects. However, we tried to counter this threat by selecting a relatively large number of projects and ensuring the diversity of the selected projects on size, lifespan, and number of contributors.
In our experiment, we only applied our experiment on the software projects that are written in Java and we did not consider projects on other Version Control Systems than GitHub. Java is one of the most popular programming languages, and GitHub is one of the most popular code-hosting platforms. 
This popularity means there are many open-source projects available for study, making the results relevant to a wide range of developers and organizations. Moreover, focusing on only one programming language can help us have a consistent feature extraction procedure that reduces the variant of the result.

Our choice of a 12-month training period and a 6-month testing period was based on trial experiments on three representative projects, where we observed that this split resulted in stable and collectable datasets. However, this configuration may not generalize optimally to all repositories. Projects with irregular activity or evolving development practices might require different window sizes for optimal performance. Future work may explore adaptive or project-specific time windows.

CoRanker relies on historical co-change information to extract features and label training data, which may limit its applicability to newly developed projects with limited history. To assess this threat, we conducted an experiment on five young projects that had only 12 months of development. We used the first 6 months to extract features and the last 6 months to label co-change relationships, and evaluated CoRanker trained on other projects. The average NDCG@5 score of CoRanker was 0.49, outperforming all baselines, including File Proximity (0.37), Code Clone (0.031), FCP2Vec (0.26), StarCoder 2–scalar (0.29), and StarCoder 2–binary (0.34). While performance is lower than on projects with longer histories, these results demonstrate that CoRanker can still provide useful predictions in new projects with limited data, thereby partially mitigating this threat.

While our experiments demonstrate the effectiveness of CoRanker in identifying co-changed methods using historical data from 150 real-world Java projects, we have not yet conducted user studies or deployed the system in active development environments. As a result, the practical impact of CoRanker on developer productivity, code quality, and workflow integration has not been empirically validated. This limits our ability to fully assess its usability and effectiveness in real-world programming contexts. We acknowledge this as a threat to external validity and plan to conduct user studies or case analyses in future work to better understand how developers interact with CoRanker and how it influences their maintenance and refactoring decisions. In addition, we did not fine-tune StarCoder 2 due to the substantial computational cost and the risk of shifting the focus of this work toward LLM engineering; fine-tuning or more advanced adaptation strategies may improve LLM-based baselines and could be explored in future work.

\textbf{Internal Threats:} The approach of employing CodeBERT to transform methods into vectors for the subsequent calculation of cosine similarity may oversimplify the complex nature of semantic similarity among methods. This simplification could potentially compromise the accuracy of our semantic similarity assessments. In RQ4, we try to find the ideal length of periods to label the dataset. The set-ups of using how much data to label the dataset we explored are not continuous and are selected by our experience. Although the results of the experiment give us many insights on the selection of labelling time period for training and testing models, it is relatively coarse-grained. Moreover, assigning relevance based on historical changes could bias the results towards frequently modified methods, regardless of their actual relevance, representing a threat to validity.

While we report comprehensive quantitative results for model comparison, we do not include a detailed qualitative analysis of prediction errors or how different model architectures (e.g., RF vs. RankNet) influence top-k recommendation quality. Such an analysis could offer additional insights into model behaviour and practical implications. We plan to conduct a more in-depth error and feature analysis in future work to complement our empirical findings.

\section{Related Work}
\label{sec:related}
In this section, we review the studies related to the usage of co-change relationships in software engineering and the studies on detecting co-change relationships in software projects.

\subsection{Co-change Relationships and Their Impact on Software Quality}

Understanding and identifying co-change relationships in software systems has been shown to have practical benefits for software maintenance and quality assurance. Prior studies have leveraged co-change information to enhance fault localization~\cite{FaultLoc1, FaultLoc2, FaultLoc3}, helping developers pinpoint fault-prone areas by considering which components have historically evolved together. Co-change relationships are also critical for guiding software changes, predicting future modifications, and improving change impact analysis~\cite{EC4ImpactAnalysis, zimmermann2005mining}. By understanding these relationships, developers can better assess which parts of the system may be affected by a change, reducing the risk of introducing defects. Therefore, improving the accuracy and prioritization of co-change detection, as we propose in this work, plays a crucial role in supporting developers and maintaining software quality.

\subsection{Identifying Co-change Relationships}

Numerous studies have aimed to detect co-change relationships from software repositories~\cite{change-impact-analysis, age-distance, historank, EConClone4, EConClone2, EConClone1}. These works typically extract change histories at the file or method level to analyze which components frequently evolve together. While useful, many of these methods face limitations in terms of granularity, noise, or lack of prioritization.

\subsubsection{Clone-based Co-change Identification}

Several studies specifically focus on co-changes between code clones~\cite{EConClone1, EConClone2, EConClone3, EConClone4}. For instance, Svajlenko et al.~\cite{EConClone1} assess how well clone detectors can identify evolutionary couplings and find that such tools are effective in uncovering some co-change patterns. Mondal et al.~\cite{EConClone4} further propose a ranking schema for evolutionary couplings in micro-clones. However, these approaches are limited to clone-related relationships and do not capture co-changes among semantically related, but non-cloned, methods. In contrast, CoRanker identifies co-change candidates without assuming they are clones.

\subsubsection{Association Rule Mining for Co-change Identification}

Association rule mining has also been applied to detect evolutionary couplings~\cite{EConClone4, zimmermann2005mining, EC4ImpactAnalysis}. These methods treat co-changes as frequent itemsets, providing insights into repeated joint modifications. However, such approaches tend to produce large, unordered sets of candidates, often leading to high false-positive rates~\cite{historank}. They also lack a mechanism to prioritize the most relevant co-changed entities. CoRanker instead frames co-change detection as a ranking task, helping developers focus on the most likely co-changing methods.

\subsubsection{Commit-level Co-change Identification}

A common strategy in prior work is to infer co-change relationships based on whether entities appear together in the same commit~\cite{EConClone4, EConClone2}. Historank~\cite{historank} builds on this idea by dynamically selecting the best ranking strategy based on performance in past revisions, and FCP2Vec~\cite{fcp2vec} proposes a lightweight technique that embeds file paths using Word2Vec and applies k-nearest neighbors to identify likely co-changed files. These methods offer improvements in efficiency or adaptability, but they are still limited in scope.

All these approaches operate at the commit level, which can introduce noise or false positives, particularly when co-changes occur only once or by coincidence. For example, two methods changed together in a single commit may not reflect a true logical dependency. Additionally, methods like FCP2Vec only utilize file path information, ignoring semantic and historical features. In contrast, CoRanker addresses these limitations by shifting the granularity from commit-level to PR level and applying a learning-to-rank model that integrates multiple signals—including semantic similarity, co-change frequency, path similarity, and authorship—into a unified ranking framework. This allows us to prioritize the most meaningful co-change relationships more accurately and effectively.

\subsection{Summary}

In summary, while prior work has provided valuable insights into co-change detection, limitations remain in handling noise, granularity, and prioritization. CoRanker addresses these gaps by integrating diverse features in a machine-learned ranking model, applied at the method level and evaluated at scale across 150 real-world Java projects.

\section{Conclusion}
\label{sec:conclusion}
To help practitioners identify the co-changed methods in a software project, we propose an approach that combines the static features of the source code and the edit history data using LtR models to recommend the most relevant co-changed methods in the PR level.

To assess CoRanker's performance, we conduct a large-scale experiment on 150 open-source projects. 
We find that the RF model is the best-performing model, reaching 0.8394 NDCG@5.
The most important feature of identifying co-change methods is the number of co-changes in history.
CoRanker significantly outperforms all six baselines by 4.7\%, 32\%, 573.5\%, 81.8\%, 41.1\%, and 59.2\%, respectively.
Finally, we find that the model with long training periods (i.e., 90 and 180 days) achieves better performance. 
The prediction power of the models starts to drop after 60 days of testing data. Therefore it is recommended to retrain the models every two months.

In the future, we aim to generalize CoRanker to support languages other than Java. 
Furthermore, we aim to integrate CoRanker to Version Control Systems by developing an Integrated Development Environment (IDE) plugin.

\bibliographystyle{ACM-Reference-Format}
\bibliography{main}

\end{document}